\documentclass[aps,prx,twocolumn,10pt,longbibliography,floatfix,superscriptaddress]{revtex4-2}
\usepackage{amsmath,amssymb,amsfonts,float,graphics,epsfig,epstopdf,color,verbatim,tabularx,multirow,appendix}
\usepackage{bm}

\usepackage{graphicx}
\usepackage{xcolor}
\usepackage{dcolumn}
\usepackage{bm}
\usepackage{amsmath}
\usepackage{xspace}
\usepackage{time}
\usepackage{comment}
\usepackage{textgreek}
\usepackage{color}
\definecolor{LinkColor}{rgb}{0.256,0.439,0.588}
\usepackage{hyperref}
\hypersetup{
colorlinks=true,
citecolor=LinkColor,
linkcolor=LinkColor,
urlcolor=LinkColor
}

\usepackage{amsmath}
\usepackage{mathtools}
\usepackage{amsfonts}
\usepackage{bm}
\usepackage{amssymb}

\usepackage{newtxmath}
\usepackage[thinlines]{easytable}
\usepackage{booktabs}

\usepackage{color}
\usepackage{xcolor}
\definecolor{darkblue}{rgb}{0.0, 0.0, 0.45}
\definecolor{darkred}{rgb}{0.64, 0.0, 0.0}
\definecolor{blue}{rgb}{0.2, 0.3, 0.85}
\definecolor{red}{rgb}{0.2, 0.3, 0.85}
\definecolor{darkgreen}{rgb}{0.0, 0.5, 0.0}
\usepackage{multibib}
\usepackage{todonotes}
\usepackage{color}
\usepackage{xcolor}
\definecolor{blue}{rgb}{0.25, 0.3, 0.95}
\definecolor{red}{rgb}{1,0,0}
\definecolor{darkgreen}{rgb}{0.0, 0.5, 0.0}

\usepackage{braket}
\usepackage{graphicx} 

\usepackage{tikz}
\usepackage{textcomp}

\def\be{\begin{equation}}
\def\ee{\end{equation}}
\def\bea{\begin{eqnarray}}
\def\eea{\end{eqnarray}}

\begin{document}

\title{Cat states carrying long-range correlations in the many-body localized phase}
\author{Nicolas Laflorencie}
\email{nicolas.laflorencie@cnrs.fr}
\affiliation{Laboratoire de Physique Th\'{e}orique, Universit\'{e} de Toulouse, CNRS, France}
\author{Jeanne Colbois}
\email{jeanne.colbois@cnrs.fr}
\affiliation{Laboratoire de Physique Th\'{e}orique, Universit\'{e} de Toulouse, CNRS, France}
\affiliation{Institut N\'eel, CNRS \& Universit\'e Grenoble Alpes, 38000 Grenoble, France}
\author{Fabien Alet}
\email{fabien.alet@cnrs.fr}
\affiliation{Laboratoire de Physique Th\'{e}orique, Universit\'{e} de Toulouse, CNRS, France}
\date{\today}
\begin{abstract}
Despite considerable efforts over the last decade, the high-energy phase diagram of the random-field Heisenberg chain still eludes our understanding, in particular the nature of the non-ergodic many-body localized (MBL) regime expected at strong disorder. In this work,
we revisit this paradigmatic model by studying the  statistics of rare atypical events of strongly correlated spin pairs traversing the entire system. They occur for unexpectedly strong disorder, i.e., in a regime where standard estimates fail to detect any instability. 
We then identify these very peculiar high-energy eigenstates, which exhibit  system-wide ${\cal{O}}(1)$ correlations, as nearly degenerate pairs of resonant cat states of the form ${\ket{\Phi}}_{\pm}\sim {\ket{\alpha_1}}\pm {\ket{\alpha_2}}$, where ${\ket{\alpha_1}}$ and ${\ket{\alpha_2}}$ are spin basis states.  We propose a simple and generic analytical description of this new class of eigenstates that exhibit system-spanning entanglement.  This analytical ansatz guides us in our search for rare hidden cat states in exponentially large many-body spectra. This also enables a systematic numerical inspection of the microscopic anatomy of these unconventional pairs, which appear in a wide range of disorder strengths. In the light of recent studies and ongoing debates on the MBL problem, our results offer new perspectives and stimulating challenges to this very active field.
\end{abstract}
\maketitle
\section{Introduction}
\label{sec:Intro}

Many-body localization (MBL) is a physical phenomenon in which isolated many-body interacting systems can avoid thermalization in the presence of strong disorder, resisting to the influence of interactions~\cite{fleishman_interactions_1980,
altshuler_quasiparticle_1997,gornyi_interacting_2005,basko_metalinsulator_2006}. A large body of literature (see reviews~\cite{sierant_many-body_2025,abanin_colloquium_2019,alet_many-body_2018}) investigated the unique properties of the MBL phase: absence of transport, memory of initial state and slow growth of quantum information after a dynamical quench, low entanglement (area law) in excited eigenstates, emergent integrability carried out by an extensive number of local integral of motion ($\ell$-bits)~\cite{serbyn_local_2013,ros_integrals_2015}, etc. These properties have been addressed with a wide variety of treatments, ranging from phenomenological approaches~\cite{huse_phenomenology_2014}, renormalization group treatments~\cite{pekker_hilbert-glass_2014,potter_universal_2015,you_entanglement_2016,thiery_microscopically_2017,morningstar_renormalization-group_2019,morningstar_many-body_2020}, more rigorous mathematical approaches~\cite{imbrie_many-body_2016,imbrie_diagonalization_2016,yin_eigenstate_2024,deroeck_absence_2024} to extensive numerics~\cite{jacquod_emergence_1997,oganesyan_localization_2007,znidaric_many-body_2008,pal_many-body_2010,bardarson_unbounded_2012,luca_ergodicity_2013,luitz_many-body_2015,gopalakrishnan_low_2015,bera_many-body_2015,serbyn_criterion_2015,lim_many-body_2016,prelovsek_density_2017,khemani_critical_2017,
doggen_many-body_2018,mace_multifractal_2019,herviou_multiscale_2019,chanda_time_2020,sierant_polynomially_2020,laflorencie_chain_2020,roy_fock-space_2021,hopjan_detecting_2021,morningstar_avalanches_2022,colbois_interaction_2024,roy_fock_2024, scoquart_role_2024,scoquart_scaling_2025}. Further, MBL physics has been probed on many different experimental platforms~\cite{schreiber_observation_2015,smith_many-body_2016,choi_exploring_2016,luschen_observation_2017,roushan_spectroscopic_2017,lukin_probing_2019,rispoli_quantum_2019,guo_observation_2021,PhysRevResearch.4.013148,leonard_probing_2023}.

One of the most important questions is how the MBL phase (or regime, see below) can be destabilized to form the ergodic thermal phase present at low disorder, and under what physical mechanisms this happens. In this very stimulating context, some recent work has also started to question the stability of the MBL phase itself in the thermodynamic limit (with arguments mainly based on finite chain numerics)~\cite{suntajs_quantum_2020,Suntajs_ergodicity_2020,sels_dynamical_2021,sirker_particle_2022,weisse_operator_2024}. This sparked some interesting debates~\cite{sierant_thouless_2020,panda_can_2020,abanin_distinguishing_2021,sierant_challenges_2022}, but also (re)opened the possibility of alternative scenarios with more than two phases:  e.g. with a non-ergodic intermediate regime between the ergodic and the “conventional” MBL phases~\cite{dumitrescu_kosterlitz-thouless_2019,weiner_slow_2019,evers_internal_2023,biroli_large-deviation_2024,colbois_statistics_2024}.
However, it should be noted that the intrinsic difficulty of the problem (out-of-equilibrium properties, exponential growth of the Hilbert space, strong disorder) makes it challenging  for numerical simulations, as only moderate finite size samples can be treated exactly.

\subsection{Instabilities}

So far, two classes of events have been argued, mostly on phenomenological grounds, to be responsible for the (thermodynamic) instability of the MBL phase: avalanches and many-body resonances (MBR). The avalanche scenario~\cite{de_roeck_stability_2017,thiery_many-body_2018} relies on the existence of rare regions of space where disorder is anomalously low, which act as nucleation seeds for thermalization of their surroundings. Under the condition that the interactions between this region and the rest of the localized particles decay on a large-enough length scale, this thermal bubble can grow and propagate as an avalanche and eventually thermalize the full system, leading to a breakdown of MBL. Since the timescales on which this process occurs are very long and difficult to capture numerically, the vast majority of work assumes that the avalanche has started and models this bubble by a thermal bath (or random matrix) that is weakly coupled to the rest of the MBL system ~\cite{luitz_how_2017,suntajs_ergodicity_2022,peacock_many-body_2023,Ha_many-body_2023,colmenarez_ergodic_2024,szoldra_catching_2024,pawlik_many-body_2024}. This approach can lead to bounds (e.g. an estimate of the disorder strength above which MBL is stable) but does not directly indicate whether avalanches are the prevailing sources of thermalization. We note that attempts at following the dynamical evolution in systems where low-disorder regions are intentionally planted lead to results possibly compatible with the avalanche scenario, yet do not demonstrate its realization in microscopic models~\cite{szoldra_catching_2024,peacock_many-body_2023}. 

Many-body resonances, on the other hand, do not require rare regions, as they correspond to the hybridization of two product-states -- which are eigenstates at infinite disorder --- by an interaction term. At lower disorder, the resonances can involve a larger number of states, leading to slightly more entangled states. A key {\it many-body} aspect highlighted in all discussions of MBR in this context~\cite{gopalakrishnan_low_2015,villalonga_eigenstates_2020,garratt_local_2021,garratt_resonant_2022,morningstar_avalanches_2022,crowley_constructive_2022,Ha_many-body_2023} is that the two product-states involved differ (from one another) extensively with the range of the resonance. In the most simple picture where only two product states are involved, MBR possibly take the form of {\it cat states} obtained by their two independent linear combinations~\cite{crowley_constructive_2022}.
Several toy models of MBR have been proposed and some of their properties studied numerically on lattice models~\cite{garratt_local_2021,garratt_resonant_2022,morningstar_avalanches_2022,Ha_many-body_2023,crowley_constructive_2022}, with Ref.~\cite{crowley_constructive_2022} highlighting that such toy resonance models can explain a wide variety of numerical finite-size observations in the intermediate to strong disorder regime of lattice models, some of which were earlier rather interpreted as evidence for MBL instability.  

The characterization of MBR is thus key to detect signs of thermalization in the controversial disorder regime. Studying the occurrence of MBR in a many-body spectrum is however expected to be an exacting task as MBR are embedded in a sea of an exponential number non-resonating MBL-like eigenstates. 
This is particularly true for {\it long-range} 
(LR) resonances which are much less frequent than short-range ones. LR resonances also only have visible effects in dynamics only on very long-time scales, essentially not accessible to numerics or experiments. These rare MBR are thus hard to probe, in spite of the facts that rare events play a crucial role in a regime where most observables display a clear MBL behavior~\cite{biroli_large-deviation_2024,colbois_statistics_2024} and that MBR are expected to be preferential channels through which avalanches can thermalize a MBL sample over larger distance~\cite{Ha_many-body_2023}.

This resulted in only a handful number of systematic, quantitative, studies of MBR in microscopic models~\cite{kjall_many-body_2018,villalonga_eigenstates_2020,morningstar_avalanches_2022,Ha_many-body_2023}. Further, the suggestion~\cite{villalonga_eigenstates_2020,villalonga_characterizing_2020,crowley_constructive_2022} that MBR take the form of {\it cat states} is a striking hypothesis that has never been neither quantitatively nor directly proven.

\subsection{Main results}
\begin{figure*}[t!]
    \centering
    \includegraphics[width=2\columnwidth]{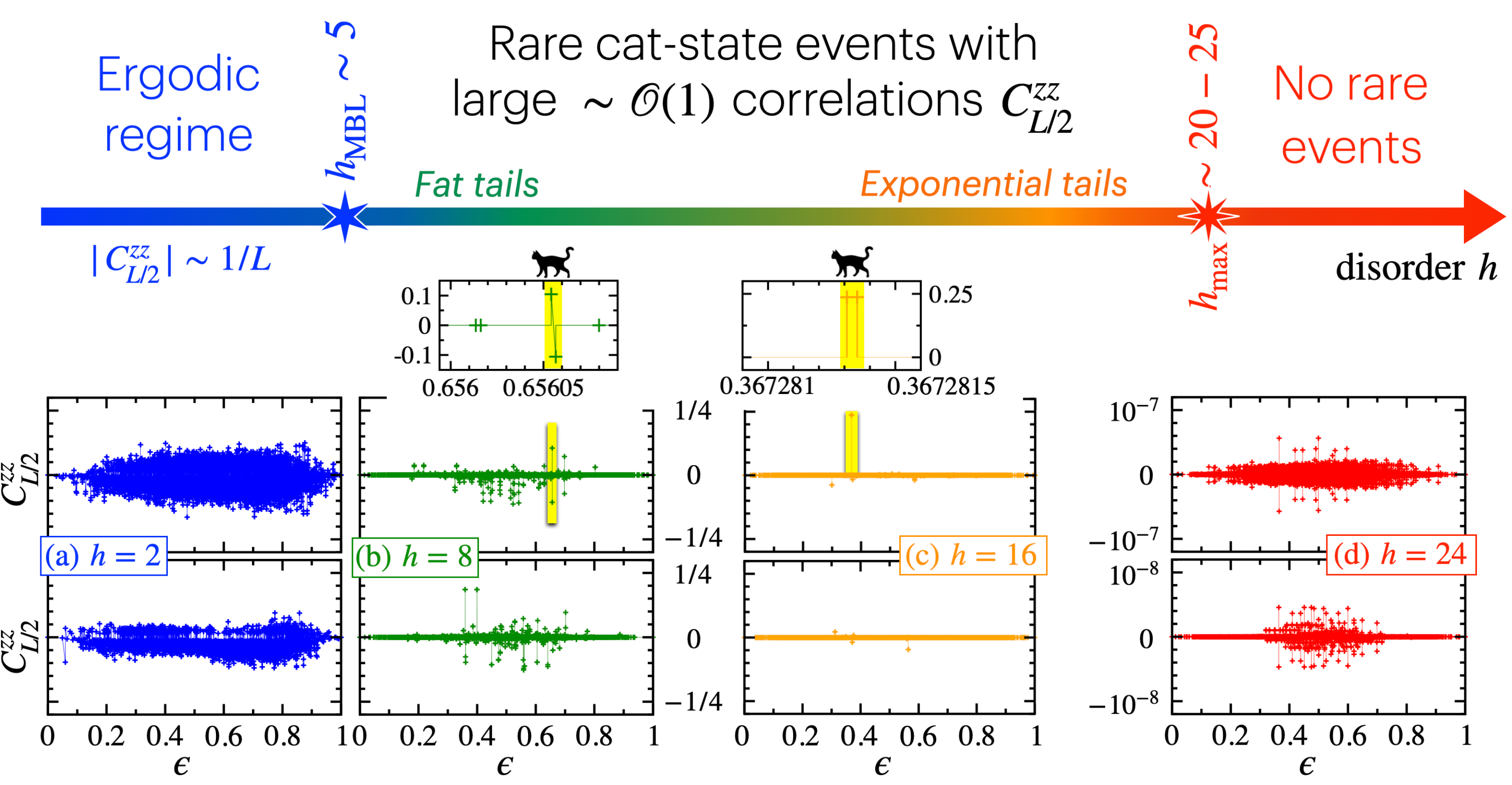}
    \caption{Overview of the different regimes of the random-field Heisenberg chain model Eq.~\eqref{eq:RFHCM}, as seen from the statistics of the midchain longitudinal correlation functions $C_{L/2}^{zz}$. The colored arrow (top) shows the main physical regimes as a function of the disorder strength $h$, and below, the energy density $\epsilon$ dependence of these correlations is plotted in the various insets, where $C_{L/2}^{zz}(\epsilon)$ is shown for all the ${\cal{N}}_{\rm H}=12870$ eigenstates~\footnote{For the sake of clarity, we show only the maximal (out of $L/2$) midchain correlations for each eigenstate instead of showing all possible $L/2$ midchain correlations.} ($L=16$ chains), for 4 representative strengths of disorder, each illustrated by two samples. Panels (a) show the ergodic behavior at $h=2$, below $h_{\rm MBL}\sim 5$ where there are no rare events, and the 2 examples shown are typical, i.e. without anomalously large correlations. In the rare event regime, panels (b) for $h=8$ and (c) for $h=16$ show the emergence of rare eigenstates across the spectrum: two pairs of cat states with anomalously large correlations are highlighted in yellow, and enlarged in the top two sub-panels where we visualize the tiny energy gaps $\Delta\epsilon$ as small as $\sim 2\cdot 10^{-6}$ (b) and $\sim 4\cdot 10^{-8}$~(c). It is also worth noting the values of the associated correlations within each pairs: (b) $C^{zz}_{L/2}\approx (0.1043,\,-0.1057)$, and (c) $C_{L/2}^{zz}\approx (0.23678,\, 0.23686)$.
    Panels (d) for $h=24$ show two typical behaviors around $h_{\rm max}\sim 20-25$ where no rare events are detected, and all eigenstates host exponentially suppressed correlations (note the very small scale on the y-axis). Between the ergodic regime (expected for $h\le h_{\rm MBL}$, where $h_{\rm MBL}\sim 5$ comes from extrapolated standard  estimates~\cite{sierant_polynomially_2020,colbois_interaction_2024}) and $h_{\rm max}\sim 20-25$ (above which {\it{all eigenstates}} exhibit short-range correlations), there is a rather broad rare-event regime that evolves continuously from fat-tail to exponential-tail behavior for the distributions $P(\ln|4C_{L/2}^{zz}|)$~\cite{colbois_statistics_2024}.}
    \label{fig:phasediag}
\end{figure*}

In this work, we propose and apply a scheme to identify LR MBR in the prototypical microscopic model of MBL for which we focus on 
rare eigenstates (at any energy) that host large connected spin-spin correlations at the longest possible distance. In the MBL phase they correspond to very unusual events, since most typical eigenstates have vanishingly small values of connected correlations due to the underlying product state structure.
These long-range correlations should be a probe for MBR at the largest scale, capturing the weakest ergodic instability within highly localized states 
at very strong disorder. Using extensive numerical simulations, we then show that they are a signature of (long-range) many-body resonances and provide a detailed microscopic picture of these unusual eigenstates. Our main results and the setup we use are summarized below, together with an outline of the paper, as well as in Fig.~\ref{fig:phasediag}, where we provide an overview of the different situations encountered upon varying the disorder strength.\\
\\
{\it Setup--- } We wish to to revisit the stability properties of the random-field Heisenberg chain model 
\begin{equation}
{\cal{H}}=\sum_i {\bf S}_i\cdot {\bf S}_{i+1}+h_i S_{i}^{z},
\label{eq:RFHCM}
\end{equation}
where each field field $h_i$ is randomly drawn from a box distribution $[-h,h]$ (with $h$ the disorder strength), in a regime $h>h_{\rm MBL}\simeq 5$ where standard localization estimates (spectral statistics, average entanglement entropy, average participation entropy etc) find an apparently well-converged MBL behavior. In Ref.~\cite{colbois_interaction_2024}, we estimate $h_{\rm MBL}\simeq 5$ from extrapolations to the {\it infinite-size} limit of these standard measures, in agreement with Ref.~\cite{sierant_polynomially_2020}.

Wishing to identify eigenstates which differ from the expected MBL behavior, our first tool of analysis is to consider eigenstates hosting {\it strong} spin-spin correlations at the longest possible distance across the chain. We define the connected pairwise correlation at distance $L/2$
\begin{equation}
\label{eq:Cdef}
    C^{\alpha \alpha}_{L/2}(i)= \langle S_i^\alpha S_{i+L/2}^\alpha \rangle- \langle S_i^{\alpha} \rangle \langle S_{i+L/2}^{\alpha} \rangle,
\end{equation}
with $\alpha=x$ or $z$ and where the expectation value is taken in a given eigenstate. The position $i$ of the first spin can be anywhere in the chain, and $i+L/2$ should be understood with the use of periodic boundary conditions. Note that $|C_{L/2}^{\alpha\alpha}|\in [0,1/4]$.  
Crucially, since eigenstates carrying very strong long-range correlations become increasingly rare as the disorder grows, one needs to compute {\it all} eigenstates to best capture rare events. This is in contrast with typical results obtain with the customary shift-invert method~\cite{luitz_many-body_2015,pietracaprina_shift-invert_2018} which allows to obtain a limited number of eigenstates (usually in the middle of the spectrum). Further, we require for our analysis a large number of disorder realizations (between $3000$ samples for $L=18$, up to $2\cdot 10^4$ for smaller chains). These two facts imply that our calculations are limited to chains of size $L\leq 18$, focusing on the largest magnetization sector of ${\cal H}$ ($S^z=0,1/2$ for even/odd $L$).\\
\\
{\it Statistics of (rare) strong correlations---} This is the main focus of Sec.~\ref{sec:stats}. By strong correlations we mean values of $|C_{L/2}^{\alpha\alpha}|$ that are larger than a certain threshold $C_\star =0.1$ or $C_\star =0.2$, thus drastically filtering the few eigenstates that satisfy such criteria, in stark contrast to the typical MBL behavior that exhibits exponentially vanishing correlation functions.

Building on our recent work~\cite{colbois_interaction_2024,colbois_statistics_2024}, where we  identified that long-range correlations can be the signature of ergodic instabilities in the regime $h>h_{\rm MBL}$, we further study the statistics and the occurrence of such atypical (strongly correlated) eigenstates over the whole many-body spectra, both in the transverse $xx$ and longitudinal $zz$ directions, and find that the longitudinal events occur in a much larger region of the phase diagram that previously reported or expected. We examine how the disorder-averaged number of eigenstates $N_{\star}^{\alpha}$ with large correlations $|C_{L/2}^{\alpha\alpha}|>C_\star$ scales with the Hilbert space size $\cal{N}_{\rm H}$. For the dominant $zz$ component we identify a crossing of the curves $N_{\star}^{z}(L)$ at very strong disorder $h_{\rm max}\sim 20-25$, above which $N_{\star}^{z}(L)\to 0$ with increasing $L$. This provides an upper bound for the critical disorder, $h_{\rm max}$, above which {\it{all eigenstates}} are expected to exhibit exponential short-range correlations, a regime that is asymptotically free of rare events.

This result is well confirmed by an extreme value study, which selects in each disorder realization a unique eigenstate, the one with the largest (in magnitude) midchain correlation ${\rm{Max}}\left|C_{L/2}^{zz}\right|$. Such an analysis shows remarkably a clear qualitative change that appears in the same region $h_{\rm max}\sim 20-25$, above which the extreme correlations eventually vanish with $L$, while below this disorder threshold they tend to their maximum magnitude $C_{\rm max}=1/4$, in contrast to previous results~\cite{morningstar_avalanches_2022} (see also Appendix~\ref{app:edges} for comparison, in particular for a discussion on the boundary conditions). 
Additional results for quantum mutual information and transverse correlations are also presented and discussed, providing more insight into the nature of the rare eigenstates that carry the strong system-wide correlations, a central issue of this work which we then discuss in Sec.~\ref{sec:LRCCS}.\\
\\
{\it{Nearly degenerate pairs of cat states---}} The very atypical class of eigenstates that carry long-range correlations, present in a wide range of disorder strengths (see Fig.~\ref{fig:phasediag}), is studied in great detail in Sec.~\ref{sec:LRCCS}. In fact, we are able to show unambiguously that they appear as {\it nearly degenerate pairs} in the many-body spectrum, taking the form of {\it cat states}, to which we devote all the rest of the study. In Sec.~\ref{sec:LRCCSToy} we first illustrate the phenomenon with a few selected examples that show very unusual features, and this leads us to build a simple but quantitatively predictive toy model description for such pairs of long-range correlated eigenstates. The {\it{catness properties}} are then explored in detail in Sec.~\ref{sec:Catness}, where we discuss and examine the most important physical properties of this class of eigenstates, which are very different from the more "standard" MBL eigenstates, i.e. those directly connected to a simple product state.

A deeper investigation of the {\it{anatomy of the cat states}} is made in Sec.~\ref{sec:CatAnatomy} where we first address the spectroscopic properties in Sec.~\ref{sec:CatSpectroscopy}, showing for example that typical cat gaps are smaller than the natural many-body level spacing. We then provide a detailed {\it microscopic portrait} of these cat states in Sec.~\ref{sec:CatMicroscopy} where we quantify and visualize the statistics of spin flips within all cat pairs by considering simple objects, namely the magnetization profiles of the cat states.\\
\\
\noindent{\it{Discussions---}} An overview of the various regimes, as seen from the statistics of the midchain
longitudinal correlation functions, is provided in Fig.~\ref{fig:phasediag} with several examples of samples showing some typical behaviors of $C_{L/2}^{zz}$ against the energy density $\epsilon$~\footnote{we use the standard notation where for a given energy level $E$,  $\epsilon=(E-E_{\rm min})/(E_{\rm max}-E_{\rm min})$ with $E_{\rm min}$ and $E_{\rm max}$ the minimal and maximal energies of the considered sample}.
Before the final conclusions, Sec.~\ref{sec:discussions} discusses the possible implications of the existence of such a rather extended rare event region hosting cat states on the phase diagram of the model, and more generally on the various debates and recent proposals aimed at refining our understanding of the MBL problem.
\section{Statistics of strong systemwide correlations}  
\label{sec:stats}
\subsection{Rare events and fat-tailed distributions of strong correlations}
Here we first elaborate and show that, even at relatively strong disorder, any  random sample may harbor some special eigenstates that show strong spin-spin correlations at long distances. By {\it{strong}} here we mean $|C_{L/2}^{\alpha\alpha}|\sim {\cal{O}}(1)$ ($\alpha=x,\,z$), corresponding to anomalous events in a regime where otherwise the typical behavior is instead exponentially suppressed with the system size $L$. These strong correlation events occur with a certain probability that we define as the following weight integrated over the full distribution of mid-chain correlators (in absolute value) ${{P}}(|C^{\alpha\alpha}_{L/2}|)$
\begin{equation}
W_{\star}^{\alpha}(L,h)=\int^{1/4}_{C_\star}{{P}}(|C^{\alpha\alpha}_{L/2}|){\rm{d}}C^{\alpha\alpha}_{L/2} ,   
\label{eq:Wzstar}
\end{equation}
which depends on disorder strength $h$, system size $L$, and the threshold value $C_\star$.  

In Ref.~\cite{colbois_statistics_2024}, we  initiated a careful exploration of rare events of strong longitudinal correlations. In particular, we identified two markedly different trends for ${{P}}(C^{zz}_{L/2})$ at strong disorder, i.e. in a regime where standard estimates based on eigenstates display localization, e.g.  spectral statistics, entanglement and participation entropies show well-converged MBL behaviors, that is typically observed beyond $h_{\rm MBL}\sim 5$~\cite{sierant_polynomially_2020,morningstar_avalanches_2022,colbois_interaction_2024}. 
However, even for $h>h_{\rm MBL}$ the distributions of long-distance correlations ${{P}}(C^{zz}_{L/2})$ are not yet converged, and exhibit fat tails~\cite{colbois_statistics_2024} on the large value side, i.e. for $C_{L/2}^{zz}\to \pm 1/4$. This regime is dominated by rare events of large ${\cal{O}}(1)$ $zz$ correlations, and their associated probability was found to decay slowly, presumably algebraically with system size 
\begin{equation}
W_{\star}^{z}(L)\sim L^{-\eta}, 
\label{eq:Wzstar_pl}    
\end{equation}
for $5\lesssim h \lesssim 8$, with $\eta\in [2,4]$. 
Then as $h$ is further increased, these fat tails get slowly suppressed, and were eventually found to disappear at larger disorder, typically above $h\sim 8-10$, where instead a faster exponential decay controlled by a finite disorder-dependent length scale  $\Lambda_z(h)$ was reported 
\begin{equation}
W_{\star}^{z}(L,h)\sim {\rm{e}}^{-{L}/{\Lambda_z(h)}}.  
\label{eq:Wzstar_exp}
\end{equation}

\subsection{Scaling with Hilbert space size}
\label{sec:scaling}
It is instructive to focus on the whole spectrum for each sample, asking what is the average number of ``atypical eigenstates'' $N_{\star}^\alpha$ that exhibit anomalously strong value of $C_{L/2}^{\alpha\alpha}$. It is  defined by
\begin{equation}
    N_{\star}^\alpha={\cal{N}}_{\rm H}\times W_{\star}^{\alpha} ,
\label{eq:Nzstar}
\end{equation}
where ${\cal{N}}_{\rm H}$ is the Hilbert space size, which grows exponentially with $L$ ($2^L$ with subdominant $1/\sqrt L$  corrections) as 
\begin{equation}
{\cal{N}}_{\rm H}\sim {\rm{e}}^{{L}/{\Lambda_{\rm H}}}\quad(1/\Lambda_{\rm H}={\ln 2}),
\label{eq:NH}
\end{equation}
$\Lambda_{\rm H}$ being the length scale associated to the many-body spectrum. 
Comparing the exponential suppression of the weight in 
Eq.~\eqref{eq:Wzstar_exp} with the exponential growth of ${\cal{N}}_{\rm H}$ in Eq.~\eqref{eq:NH}, we can anticipate the possible existence of (at least) two main regimes characterizing different scalings with ${\cal{N}}_{\rm H}$ of the number of eigenstates $N_{\star}^{z}$ that exhibit large $zz$ correlations.

(i) If $\Lambda_z(h)>\Lambda_{\rm H}$, $N_{\star}^z(L)$ will grow exponentially with $L$, but only as a vanishing fraction of the total Hilbert space size: $N_{\star}^z\sim {\cal{N}}_{\rm H}^{d_\star}$, with an "effective dimension" $d_\star=1-{\Lambda_{\rm H}}/{\Lambda_z}$.

(ii) If on the other hand $\Lambda_z(h)< \Lambda_{\rm H}$, the above effective dimension will change sign $d_\star < 0$, corresponding to $N_{\star}^z\to 0$ at large $L$, i.e. an absence of rare events.

(iii) In addition to these two distinct behaviors, there is a third possibility, for the particular case of the fat-tail power-law regime Eq.~\eqref{eq:Wzstar_pl}. This would morally correspond to having a length scale $\Lambda_z$ that effectively grows with $L$, and thus a number of long-range correlated eigenstates $N_{\star}^z(L)$ that is expected to scale very fast, almost as fast as the Hilbert space dimension. However, as we will see below, the distinction between cases (i) and (iii) is not easy to capture accurately in our numerical data.
\subsection{Statistics of rare events: numerical results}
\label{sec:statnum}
\subsubsection{Longitudinal correlations $C_{L/2}^{zz}$}
  \begin{figure}[t!]
    \centering
    \includegraphics[width=\columnwidth]{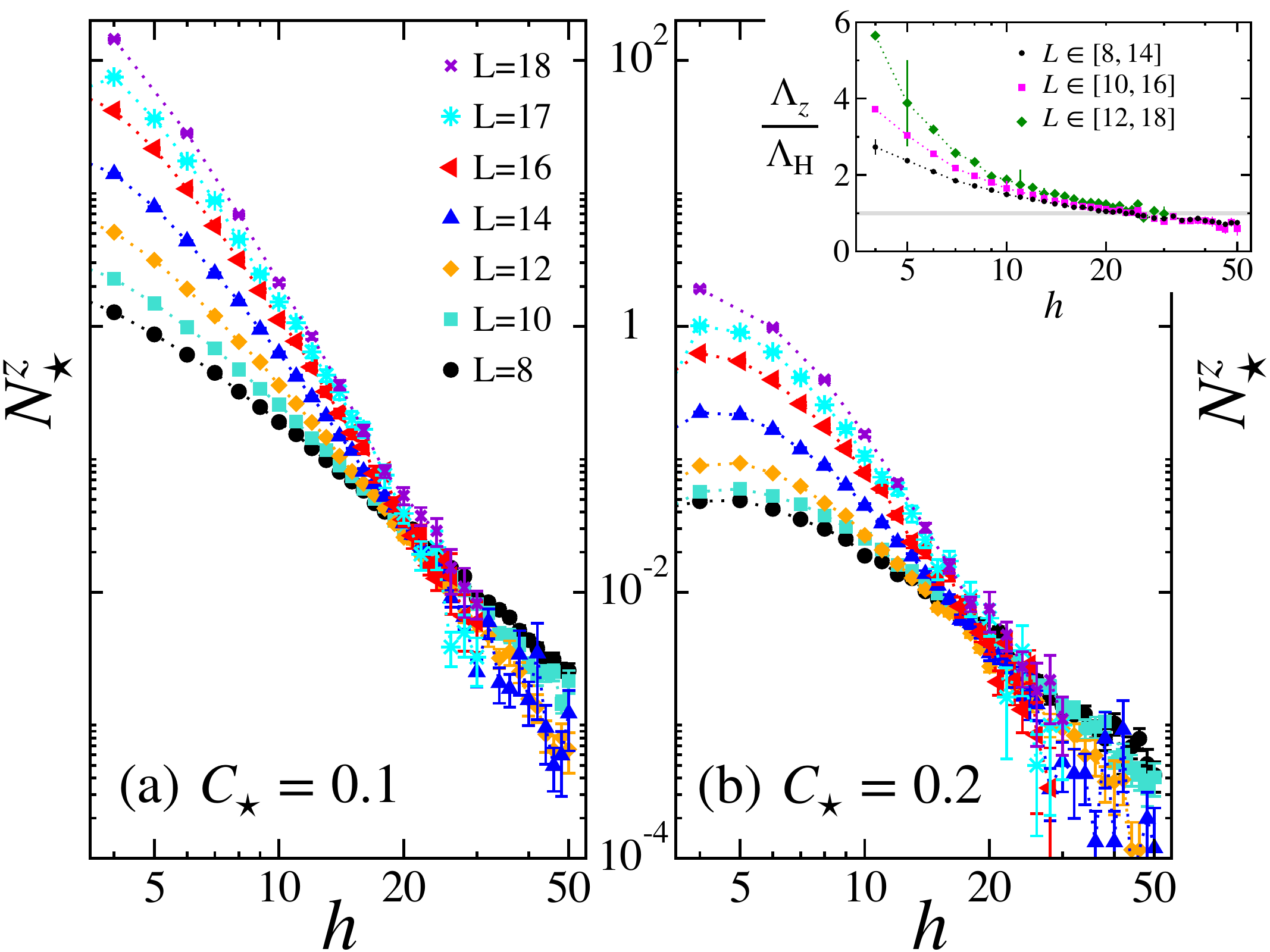}
    \caption{Average number of eigenstates (per sample) $N_\star^z$ having strong systemwide $zz$ correlations, i.e. mid-chain correlations such that  $|C_{L/2}^{zz}|\ge C_{\star}$, with (a) $C_\star=0.1$ and (b) $C_\star=0.2$.
    Plotted against disorder strength $h$, the different curves $N_\star^z(L,h)$ for chain lengths $L\in [8,\,18]$, exhibit a clear crossing for $h\sim 20$. Inset shows how the length scale $\Lambda_z$ for $C_\star=0.1$ (rescaled by the spectral length $\Lambda_{\rm H}=\frac{1}{\ln 2}$) decays with $h$ for various fitting windows indicated on the plot. The line $\Lambda_z/\Lambda_{\rm H}=1$ is crossed for $h\sim 20$. Finite size effects become significant below $h\sim 10$.}
    \label{fig:Nz}
\end{figure}

We first show in the main panels of Fig.~\ref{fig:Nz} the behavior of $N_{\star}^{z}$ (associated to longitudinal correlators) for $C_\star=0.1,\, 0.2$ as a function of disorder strength $h$ and for increasing system sizes ($L=8,\,10,\,\cdots,\,17,\,18$).
Two different scalings can be clearly identified. Below $h\sim 20$, $N_{\star}^{z}$ grows with system size, while for larger disorder, it scales to zero. This indicates that for sufficiently large disorder, events with strong system-wide correlations do not occur in the thermodynamic limit. Clearly, at such large values of disorder $h \gtrsim 20$, the extreme rarity of these events makes their detection very challenging, and a huge sampling effort is required to obtain decent statistics, which is signaled by a stronger noise to signal ratio in this regime (note the log-log scales of the main panels). This effect is amplified for larger values of the threshold $C_\star$ (see right panel for $C_\star=0.2$), but nevertheless the same behavior is found, with a threshold that also takes place in the region $h\sim 20$.

The disorder-dependence of the associated length scale $\Lambda_z(h)$ (rescaled by the spectral length $\Lambda_{\rm H}$) is shown in the inset of Fig.~\ref{fig:Nz} (for $C_\star=0.1$). 
As previously anticipated, at strong disorder, typically beyond $h\sim 20$, we observe a well converged behavior with $\Lambda_z/\Lambda_{\rm H}<1$ which means that $N_{\star}^{z}\to 0$ with $L$, corresponding to an asymptotic absence of rare events. On the other side, the regime (i) where $\Lambda_z>\Lambda_{\rm H}$ is perfectly observed, but one also notice strong finite-size effects in the regime $h\lesssim 10$, which is consistent with previous results on the fat-tail regime (see Ref.~\cite{colbois_statistics_2024} and also a more detailed discussion in Appendix~\ref{app:twopoint}).

\subsubsection{Transverse correlations $C_{L/2}^{xx}$}

 \begin{figure}[t!]
    \centering
   \includegraphics[width=\columnwidth]{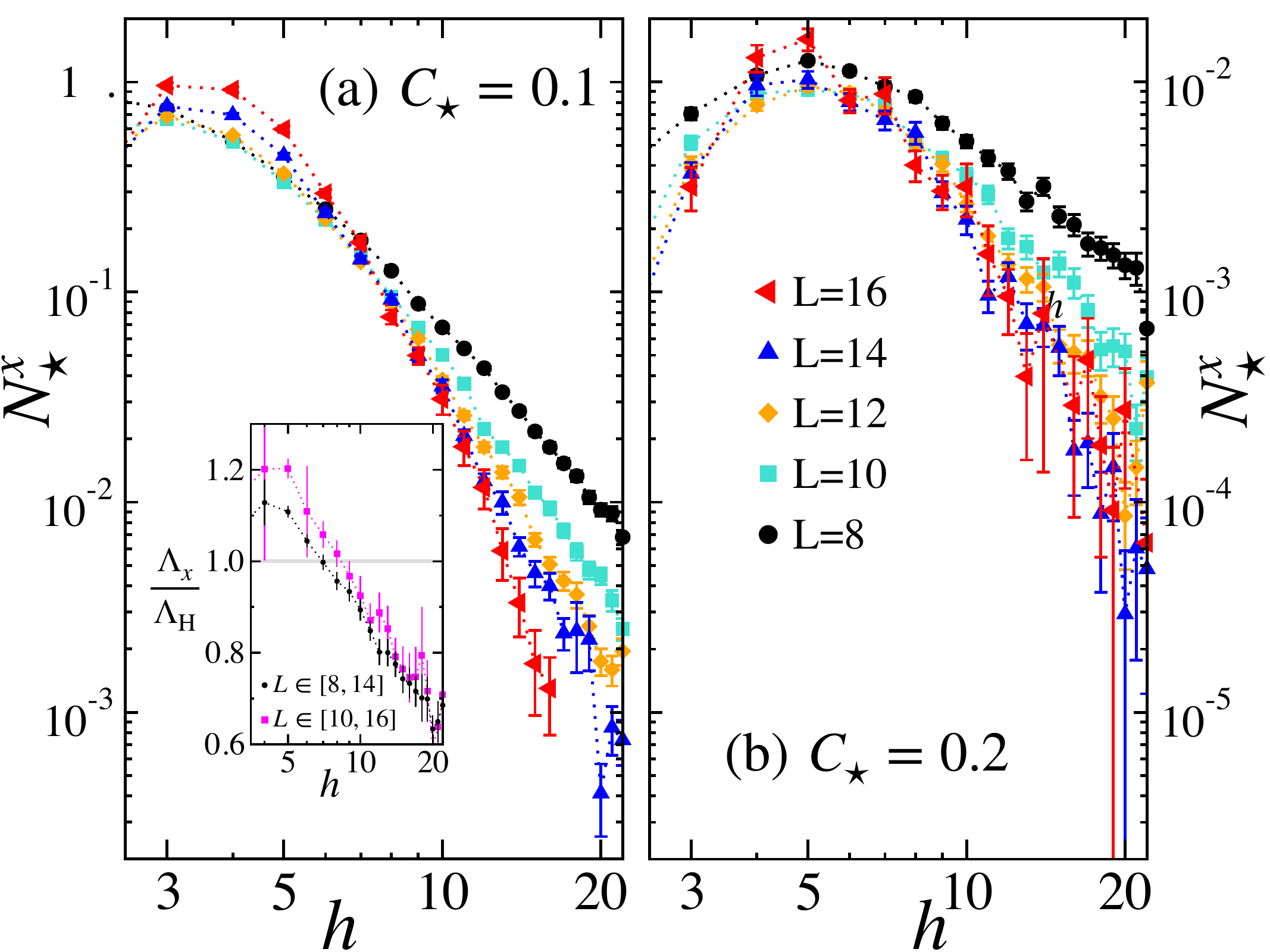}
    \caption{Same as Fig.~\ref{fig:Nz} for the transverse component $N^*_x$ with (a) $C_\star=0.1$ and (b) $C_\star=0.2$. The inset shows length scale $\Lambda_x$ for $C_\star=0.1$, rescaled by the spectral length $\Lambda_{\rm H}=\frac{1}{\ln 2}$. The line $\Lambda_z/\Lambda_{\rm H}=1$ is crossed for $h\sim 10$, but with significant finite-size effects.}
    \label{fig:Nx_and_Lambdax}
\end{figure}
In contrast to the $zz$ component, the number of large transverse correlation events is much smaller, see Fig.~\ref{fig:Nx_and_Lambdax}. Since it is numerically more demanding to compute these off-diagonal objects, we have limited these calculations to $L\le 16$, but this is clearly sufficient to see the strong differences between $N_{\star}^{z}$ and $N_{\star}^{x}$, the latter hardly reaching unity (i.e., less than 1 eigenstate with large $C^{xx}$ on average) even for the lowest disorder, larger sizes and smaller threshold.
For the larger threshold $C_\star=0.2$ there are almost no events (more precisely, on average less than 1 event out of 100 disorder samples), resulting in very poor statistics.

As for the longitudinal correlators, a crossing of the curves $N_{\star}^{x}(L)$ can also be identified, but with a strong finite-size drift.  An exact value for the crossing point is hard to determine, but it seems clear that it occurs for significantly {\it lower} disorder strengths (compared to $N_{\star}^{z}$), typically around $h\sim 8-10$ for $C_\star=0.1$ (but practically impossible to detect for $C_\star=0.2$ due to too low statistics).

These results indicate that events of very large $xx$ correlators are much rarer than those of large $C^{zz}_{L/2}$, and seem to exist in the thermodynamic limit only in a very limited range of disorder. Similar conclusions will be reached for the extreme eigenstate statistics, which we examine in the next subsection.

\subsection{Extreme correlations across the spectrum}
\label{sec:ExtremeCorr}
\subsubsection{Setup}
To address the ``most delocalized'' events traversing the chain, we systematically collect the largest system-wide correlations, i.e. at distance $L/2$, in {\it{every}} sample, hosted by ${\it a single}$ eigenstate in the {\it{whole set}} of ${\cal{N}}_{\rm H}$ eigenstates. We discuss the two correlators $C_{L/2}^{xx}$ and $C_{L/2}^{xx}$, as well as the pairwise quantum mutual information (QMI). This extreme value calculation requires a rather large averaging over the disorder: we use between $2\cdot 10^4$ for the smallest systems, $6\cdot 10^3$ for $L=16$, and $3\cdot 10^3$ samples for $L=18$. In the disorder averaging process, we compute and present below {\it{the typical values}}, which, as often, display less fluctuations.

\subsubsection{Diagonal correlations}

    \begin{figure}[b!]
    \centering
   \includegraphics[width=\columnwidth]{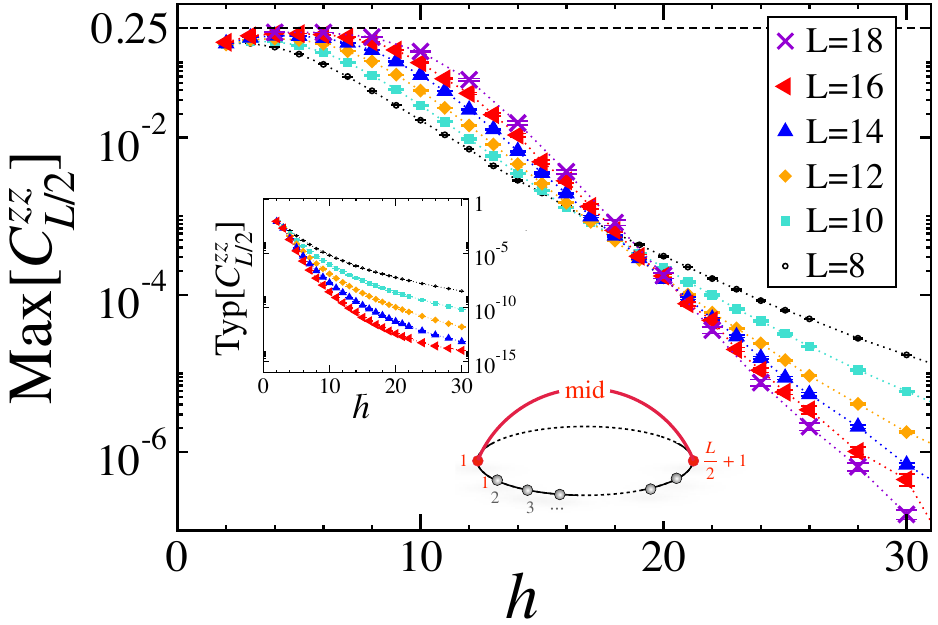}
    \caption{Maximal value of the long-distance longitudinal correlation $C_{L/2}^{zz}$ per sample. The maximum is computed for each random sample over all the ${\cal{N}}_{\rm H}$ eigenstates (and the $L/2$ possible pairs), and then averaged over a large number of samples (typically $\sim 2\cdot 10^4$ for $L\le 14$, $\sim 10^4$ for $L=16$, and $3\cdot 10^3$ for $L=18$). The inset shows the typical average computed over {\it all} the eigenstates for comparison: note the very different scale on the y-axis.} 
    \label{fig:max_Czz}
\end{figure}
We first discuss the dominant systemwide response, that is the $zz$ correlation function at mid-chain $C_{L/2}^{zz}$, shown in 
Fig.~\ref{fig:max_Czz}. A few comments are in order:

(i) For large disorder strengths, above $h\sim 20$, even the most strongly correlated eigenstate in each sample --- the one with the strongest $zz$ correlation across the system --- is found to have a vanishing correlator $C_{L/2}^{zz}$ (among the $L/2\times {\cal{N}}_{\rm H}$ possible values) as $L$ increases. Therefore, this threshold $h_{\rm max}\sim 20-25$ can be interpreted as an upper bound for the MBL transition, in remarkable agreement with the analysis of $N_{\star}^{z}\to 0$ in Fig.~\ref{fig:Nz}, which also reaches similar conclusions for $h_{\rm max}$.

(ii) Below $h_{\rm max}$, the largest $zz$ correlations increase with $L$ and are expected to saturate, presumably at the maximum possible value $C_{\rm max}=1/4$. 

(iii) For weak disorder, although our study is not focused on the ergodic side, our data show a tendency for $C_{\rm max}^{zz}$ to decay slowly below $h\sim 4$, but with unclear finite size effects~\footnote{See for instance Ref.~\cite{haque_entanglement_2022} where deviations from ETH are discussed (I. Khaymovich, private communication)}.

It should be noted that this extreme value analysis does not give much information about the fat tail regime, which is roughly expected in the range $h \sim(5-10)$, because these extreme eigenstates certainly have similar properties throughout the rare event region.

\subsubsection{Transverse correlations and quantum mutual information}
    \begin{figure}[b!]
    \centering
   \includegraphics[width=\columnwidth,clip]{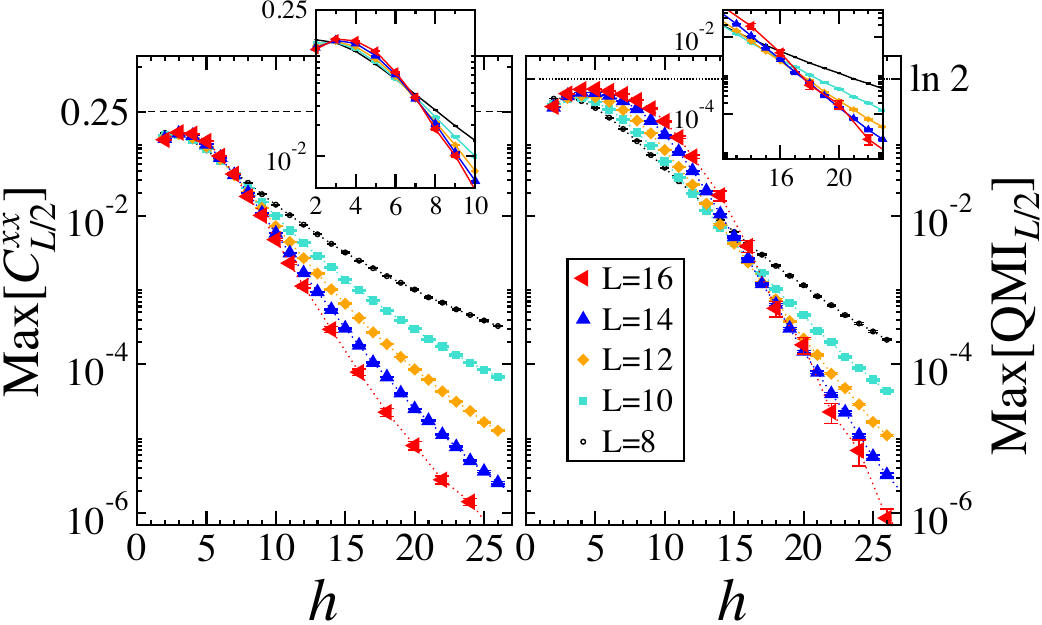}
    \caption{Same as Fig.~\ref{fig:max_Czz} for $C_{L/2}^{xx}$ (left) and QMI$_{L/2}$ (right). The insets show zooms on the crossing regions.}
    \label{fig:max_Cxx_and_QMI}
\end{figure}

One can repeat the same extreme value analysis for the transverse correlation functions $C_{L/2}^{xx}$. The results are shown in Fig.~\ref{fig:max_Cxx_and_QMI} (left) where again a crossing is observed, but for a significantly smaller value of the disorder strength, about $h\sim 8$, a result compatible with the $N_{\star}^{x}$ analysis shown in Fig.~\ref{fig:Nx_and_Lambdax}.

Fig.~\ref{fig:max_Cxx_and_QMI} (right) shows the related and complementary two-body object: the 2-site QMI (see Appendix~\ref{app:qmi}), defined by
\be
{\rm QMI}_{ij}=S^{(1)}_{i}+S^{(1)}_{j}-S^{(2)}_{ij},
\label{eq:QMI}
\ee
where $S^{(1)}_{i}$ is the single site ($i$) entanglement entropy (EE) and $S^{(2)}_{ij}$ the EE of the two sites $(i,j)$ subsystem. 
QMI$_{ij}$ takes values in the range $[0,2\ln 2]$, and the connection of with pairwise correlators is detailed in Appendix~\ref{app:qmi} along with its computation for simple cases.

QMI has been argued~\cite{tomasi_quantum_2017,morningstar_avalanches_2022} to be a relevant witness to MBL physics. Here, focusing on its extreme statistics, we show in the right panel of Fig.~\ref{fig:max_Cxx_and_QMI} its maximal value computed for each sample over all the eigenstates and all the $L/2$ possible maximal distances $|i-j|=L/2$ for PBC. 
Interestingly, the different curves Max(QMI$_{L/2}$) cross roughly at the same value as the one observed for $C_{L/2}^{zz}$, i.e. for $h_{\rm max}\sim 20-25$, although with a slightly more pronounced finite-size drift of the crossing points. 

The quantum structure of the system-wide resonances thus appear to be dominated by strong $zz$ correlations in most of the regime, yielding for example a QMI that saturates towards a maximum value of $\ln 2$, while in principle one could also expect to observe a ${\rm{QMI}}_{\rm max}=\ln 4$. This fact is discussed further below in section ~\ref{sec:LRCCSToy}, where we propose a toy model description for these extreme events, and for which we are able to compute some observables, such as the QMI, as presented in Sec.~\ref{sec:predictiontoystates}.

\subsubsection{Remarks on the choice of boundary conditions}
\label{sec:boundary}

It is interesting to comment on the related analysis by Morningstar {\it{et al.}}~\cite{morningstar_avalanches_2022}, where they also study the extreme values (among all eigenstates) of the two-site QMI. In contrast to our analysis, they focus on the boundary spins of {\it open} chains and find a crossing for significantly smaller disorder, typically for $h\sim 8$, a value we also confirm with this OBC setup (see Appendix~\ref{app:edges}).
It is important to observe that for strong disorder, the two (left and right) boundary spins of a chain with OBCs are atypical sites because they effectively experience random fields that appear stronger, when compared to the other terms in the Hamiltonian (kinetic and interaction terms) that are effectively reduced by a factor of two due to the locally smaller connectivity. Therefore, these edge spins will appear ``more localized'' than the bulk, and this can be readily observed numerically, as detailed in Appendix~\ref{app:edges} by considering edge magnetization or end-end correlators. 
This boundary effect affects the finite-size results sufficiently to lead to an underestimation of the system-wide resonance landmark proposed in   Ref.~\cite{morningstar_avalanches_2022}, due to this choice of open chains. The new estimate obtained with periodic chains (see Appendix~\ref{app:edges}) is pushed to a larger value, that appears close to another landmark, namely the avalanche threshold estimate, provided in Ref.~\cite{morningstar_avalanches_2022}.

\section{Cat states carrying long-range correlations}
\label{sec:LRCCS}
Given the accumulated evidence for eigenstates with unexpectedly strong correlations presented in Sec.~\ref{sec:stats}, it is natural to ask what is the nature of these eigenstates, whether they have some specific structure or exhibit unusual properties ,beyond their mere statistical occurrence. This section aims at providing elements of answers to these questions. 

We have hinted in Sec.~\ref{sec:Intro} and Fig.~\ref{fig:phasediag} at the fact that in the strong disorder regime, large correlations at long distances seem to appear in {\it pairs}, suggestive of cat states, throughout the many-body spectrum. This will first be illustrated in  Sec.~\ref{sec:LRCCSToy} by two concrete examples, which will then allow us to build a toy model of the cat states that will be helpful to characterize their microscopic properties. The demonstration of the existence of cat states in a large extended region of the phase diagram will be done in Sec.~\ref{sec:Catness} by confronting the toy cat states with the true many-body eigenstates with large correlations. We will further inspect the anatomy (spectral properties, microscopic structure such as magnetization profile) of this family of long-range correlated cat states in Sec.~\ref{sec:CatAnatomy}. 
    \begin{figure*}
            \centering
   \includegraphics[width=1.5\columnwidth]{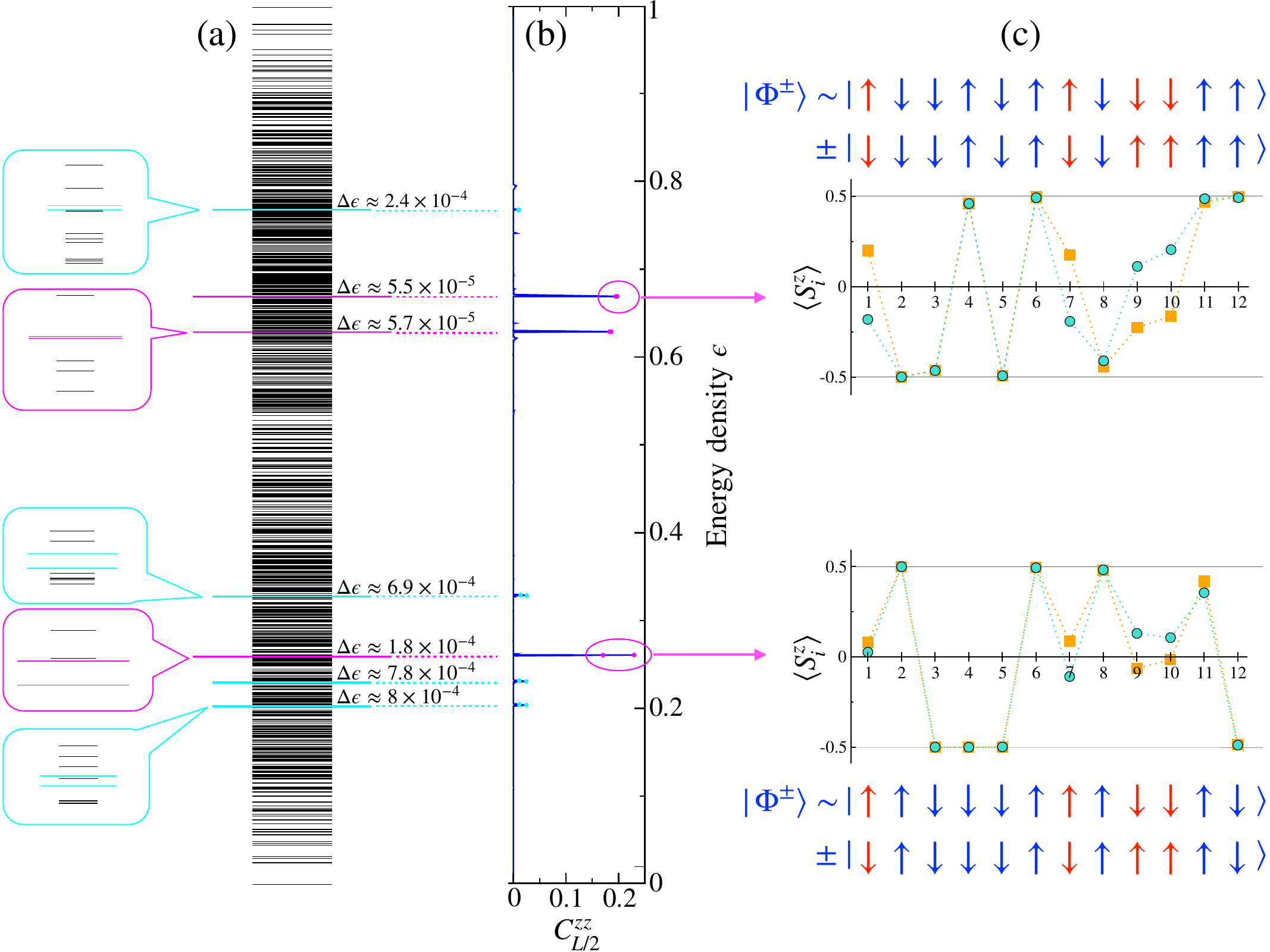}
    \caption{\emph{Sample 1} -- Illustrative example of a single realization ($L=12$, $h=20$) exhibiting 7 pairs of nearly degenerate resonant "cat states". (a) Full spectrum with the 7 pairs highlighted. The corresponding gaps are smaller or comparable to the typical level spacing $\approx 4.57\cdot 10^{-4}$ for $h = 20$, $L=12$. (b) Longitudinal correlations $C_{L/2}^{zz}$ in the middle of the spectrum $\epsilon\in [0,1]$. Two thresholds are used: $C_\star=0.15$ (magenta) and $C_\star=0.01$ (cyan). (c) Magnetization profiles for the two highlighted pairs, and sketch of the associated cat states. In both cases, sites (1) and (7) are the ones involved in the strong long-range correlations.} 
    \label{fig:spectrum}
    \end{figure*}

\subsection{Toy description of long-range correlated eigenstates}
\label{sec:LRCCSToy}

\subsubsection{Examples of atypical long-range-correlated eigenstate(s)}
Here, we present two examples of samples that exhibit very remarkable features, and explore in detail the eigenstates that emerge from our numerical observation of strongly correlated pairs of sites at distance $L/2$.  These features will in turn help to refine our search for cat states.

{\it Sample 1 ($L=12$, $h=20$)---}~Fig.~\ref{fig:spectrum} gives a closer look at the first example of a short ($L=12$) but strongly disordered ($h=20$) chain. It has about 14 eigenstates (out of 924) that show large correlations, i.e. above a certain threshold, fixed for this specific illustrative example to be $C_\star=0.15$ (magenta) and $C_\star=0.01$ (cyan), see Fig.~\ref{fig:spectrum} (b) for the $\epsilon$-dependence of $C_{L/2}^{zz}$. For this particular sample, the strong midchain correlations appear between sites $1$ and $7$ for all these eigenstates. Remarkably, they come by {\it pairs}, embedded in the eigenspectrum, with suprisingly small energy splittings, as clearly shown in Fig.~\ref{fig:spectrum} (a).

The magnetization profiles $\langle S_i^z\rangle$, shown in panels (c) for the two pairs of eigenstates with the largest $C^{zz}_{L/2}$, display unique features. In each pair, the profiles are very similar for the two partners, with some spins almost frozen to their maximal magnetization $|\langle S_i^z\rangle |\simeq 1/2$, while the others appear as {\it resonating} as they have much smaller expectation values (closer to $|S_i^z|\simeq 0$) that are opposite within the pair, like for instance the strongly correlated spins $(i,i+L/2)=(1,7)$ which belong to such resonating sites.
This suggests a cartoon representation for each pairs of eigenstates as {\it cat states} which matches the strongly correlated sites as well as the $\langle S_i^z\rangle$ data, as illustrated above the magnetization profiles in Fig.~\ref{fig:spectrum} (c). 

\begin{figure}[hb!]
    \centering
    \includegraphics[width=\columnwidth]{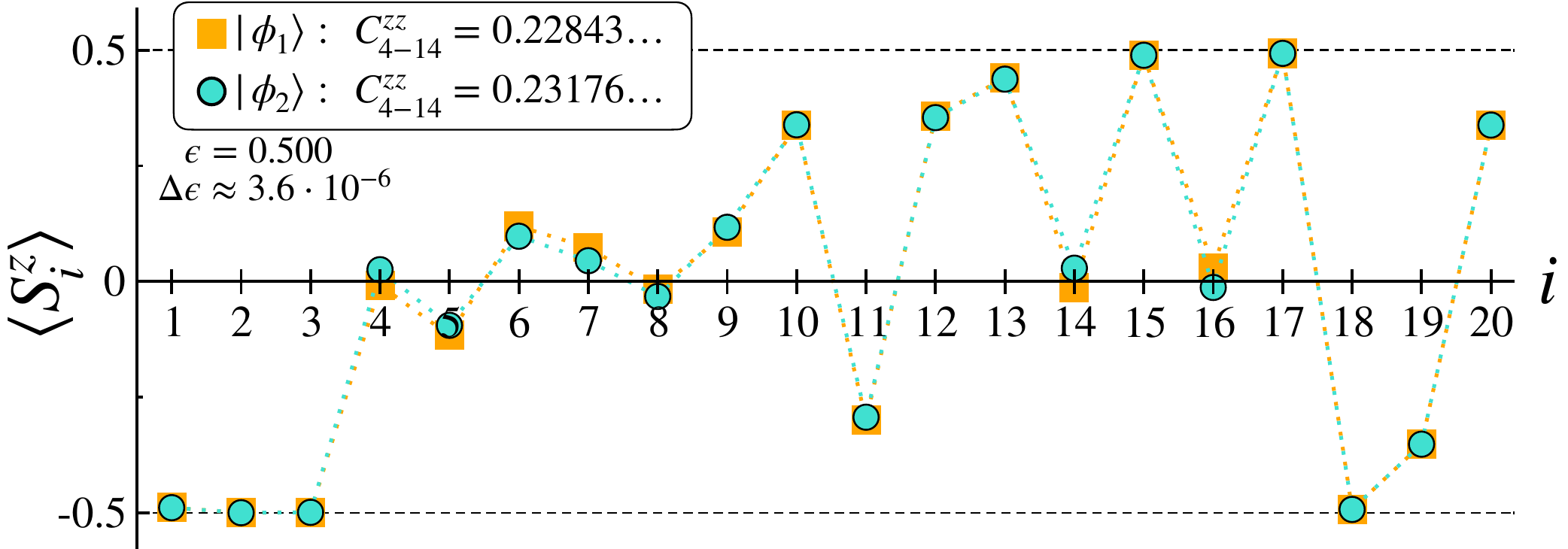}
    \caption{\emph{Sample 2} -- Illustrative example of a single realization for a PBC chain of length $L=20$ for (not too) strong disorder $h=9$ showing a pair of nearly degenerate ($\Delta\epsilon\approx 3.6\times 10^{-6}$) mid-spectrum ($\epsilon=0.5$) eigenstates $\ket {\phi_1}$ and $\ket{\phi_2}$. Strongly correlated sites ($4-14$) are detected at the largest distance ($r=10$), with $C_{L/2}^{zz}\approx 0.228$ for $\ket {\phi_1}$ and $C_{L/2}^{zz}\approx 0.232$ for $\ket{\phi_2}$ (see also Appendix~\ref{app:examplemore}). The two magnetization profiles ${\bra {\phi_1}} S_i^z {\ket{\phi_1}}$ and ${\bra {\phi_2}} S_i^z {\ket{\phi_2}}$ are remarkably almost identical.}
\label{fig:sample_L20_h9} 
\end{figure}

{\it Sample 2 ($L=20$, $h=9$)---}~The second example for a larger $L=20$ chain is shown in Fig.~\ref{fig:sample_L20_h9} for two eigenstates having strong midchain correlations between the sites $(i,i+L/2)=(4,14)$. Their magnetization profiles are found to be almost identical, but unlike the first example, here the weaker disorder $h=9$ causes the spins between the strongly correlated sites to be slightly less polarized. However, again the states appear to be nearly degenerate over the exponentially large many-body spectrum, with a tiny normalized gap $\Delta\epsilon \approx 3.6\times 10^{-6}$, slightly smaller (but typically of the same order) than the natural mid-spectrum spacing for $L=20$: $1/{\cal{N}}_{\rm H}\sim 5.4\cdot 10^{-6}$.

Based on these examples, and the possibility to think of these strongly correlated eigenstates as cat states, in the next subsection we present candidate toy cat states for which we can compute specific properties that will then be compared to real many-body calculations.

\subsubsection{Toy cat states for pairs of eigenstates with strong correlations}
\label{sec:toystates}
We propose to describe cat states hosting strong correlations using simple ansatz wave functions with a single parameter $\theta$. We start with the simplest case of two-site Bell-type states (i), which we then generalize to $L$-site cat states (ii).\\
{\it{(i) Two-site Bell's states---}} Following the standard notations of Bell's pairs, we introduce two families of (normalized) states ${\ket{\Psi}}$ and ${\ket{\Phi}}$, defined for two resonant sites $(ij)$ by
\begin{equation}
\label{eq:Psi}
\begin{array}{l}
{\ket{\Psi_{ij}^{+}}}=\cos\left(\frac{\theta}{2}\right){\ket{\uparrow_i\downarrow_j}}+\sin\left(\frac{\theta}{2}\right){\ket{\downarrow_i\uparrow_j}}\\
{\ket{\Psi_{ij}^{-}}}=\sin\left(\frac{\theta}{2}\right){\ket{\uparrow_i\downarrow_j}}-\cos\left(\frac{\theta}{2}\right){\ket{\downarrow_i\uparrow_j}},
\end{array}
\end{equation}
which describe $S^z_{\rm tot}= 0$ states, and 
\begin{equation}
\label{eq:Phi}
\begin{array}{l}
{\ket{\Phi_{ij}^{+}}}=\cos\left(\frac{\theta}{2}\right){\ket{\uparrow_i\uparrow_j}}+\sin\left(\frac{\theta}{2}\right){\ket{\downarrow_i\downarrow_j}}\\
{\ket{\Phi_{ij}^{-}}}=\sin\left(\frac{\theta}{2}\right){\ket{\uparrow_i\uparrow_j}}-\cos\left(\frac{\theta}{2}\right){\ket{\downarrow_i\downarrow_j}},
\end{array}
\end{equation}
if the total magnetization is not conserved. Within each of these 2 families, eigenstate pairs $(\pm)$ are orthogonal to each other ${\braket{\Phi^- |\Phi^+}}={\braket{\Psi^- |\Psi^+}}=0$.
The angle $\theta\in [0,\pi]$ allows a continuous description, going  from simple product states when $\theta$ is close to $\pi$ or 0, to maximally entangled cat states, when $\theta\approx \pi/2$. Below we aim at discussing several physical quantities for the whole range $0\le \theta\le \pi$, keeping in mind that we will be mostly interested in the true cat regime, i.e. for $\theta\approx \pi/2$.\\

One- and two-site observables can be easily computed within such simple ansatz wave functions, as given in Tab.~\ref{tab:catstates}. As expected, good cat states ($\theta\approx \pi/2$) give small local spin polarizations $\sim \cos\theta$ and large two-site $zz$ correlations $\propto \sin^2\theta$ for both families, while  $\Psi$ and $\Phi$ give very different results for the transverse correlations.

\begin{table}[h!]
\begin{tabular}{c|c|c}
        & $\ket{\Phi^{\pm}_{ij}}$&  ${\ket {\Psi^\pm_{ij}}}$  \\ 
        \hline
$2\langle S_i^z \rangle$ & $\pm \cos\theta$& $\pm \cos\theta$ \\
$2\langle S_j^z \rangle$ & $\pm \cos\theta$& $\mp \cos\theta$ \\ \hline
$4 C_{ij}^{zz}$ & $\sin^2 \theta$& $-\sin^2 \theta$ \\ 
$4 C_{ij}^{xx}$ & 0 & $\pm \sin\theta$ \\ \hline
\hline
\end{tabular}
\caption{Expectation values of local magnetizations and two-point correlations for cat states of  the Bell-type $\ket{ \Psi^{\pm}_{ij}}$ Eq.~\eqref{eq:Psi} for $S^z_{\rm tot}=0$, and $\ket{ \Phi^{\pm}}$ Eq.~\eqref{eq:Phi} for $S^z_{\rm tot}\neq  0$.}
\label{tab:catstates}
\end{table}

Our numerical results (Sec.~\ref{sec:statnum} and Sec.~\ref{sec:ExtremeCorr}) clearly show that the dominant correlations occur along the longitudinal channel ($zz$), while in most of the rare-event regime, the transverse ($xx$) correlations are very small. This strongly suggests that the ${{\Psi}}$ form may not be the dominant type of cat states.
Moreover, it is crucial to notice that for models conserving the total magnetization, such as the random-field Heisenberg Hamiltonian, states of type $\Phi$ are not sufficient to describe a pair of resonating sites $(ij)$, and one needs to allow for more fluctuating spins in the ansatz, as we discuss now.\\
\\
{\it{(ii) More general cat states---}} Considering now a larger system with an even number of $L$ spins (generalization to odd $L$ is straightforward), a minimal ansatz wave function of the form $\Phi$ of Eq.~\eqref{eq:Phi}, that conserves the total magnetization, and describe two-site resonances necessarily requires (at least) 2 additional fluctuating sites, leading to the minimal 4-site resonant states of the form
\begin{equation}
\label{eq:Phi4}
\begin{array}{l}
{\ket{{\Phi_{4}^{+}}}}=\Bigl(\cos\left(\frac{\theta}{2}\right){\ket{\uparrow_i\uparrow_j\downarrow_{k}\downarrow_{\ell}}}
+\sin\left(\frac{\theta}{2}\right){\ket{\downarrow_i\downarrow_j\uparrow_{k}\uparrow_{\ell}}}\Bigr)\otimes {\ket{\varphi_{L-4}}}\\
{\ket{{\Phi_{4}^{-}}}}=\Bigl(\sin\left(\frac{\theta}{2}\right){\ket{\uparrow_i\uparrow_j\downarrow_{k}\downarrow_{\ell}}}
-\cos\left(\frac{\theta}{2}\right){\ket{\downarrow_i\downarrow_j\uparrow_{k}\uparrow_{\ell}}}\Bigr)\otimes {\ket{\varphi_{L-4}}},
\end{array}
\end{equation}
where $i,\,j,\,k,\,\ell$ are 4 fluctuating sites, and ${\ket{\varphi_{L-4}}}$ is a (zero total magnetization) spin-basis state of the other remaining $L-4$ sites. This construction can be generalized to $2p$ fluctuating sites (that can be anywhere in the system, not necessarily neighbors):
\begin{equation}
\label{eq:Phi2p}
\begin{array}{l}
{\ket{{\Phi_{2p}^{+}}}}=\left(\cos\left(\frac{\theta}{2}\right){\ket{\varphi_{2p}}}
+\sin\left(\frac{\theta}{2}\right){\overline{{\ket{\varphi_{2p}}}}}\right)\otimes {\ket{\varphi_{L-2p}}}\\
{\ket{{\Phi_{2p}^{-}}}}=\left(\sin\left(\frac{\theta}{2}\right){\ket{\varphi_{2p}}}
-\cos\left(\frac{\theta}{2}\right){\overline{{\ket{\varphi_{2p}}}}}\right)\otimes {\ket{\varphi_{L-2p}}},
\end{array}
\end{equation}
where the ${\ket{\varphi_{2p}}}$ are $2p$ site product states of zero total magnetization, for instance of the form ${\ket{\varphi_{2p}}}={\ket{\uparrow\uparrow\uparrow\downarrow\downarrow\downarrow}}$, with ${\overline{{\ket{\varphi_{2p}}}}}={\ket{\downarrow\downarrow\downarrow\uparrow\uparrow\uparrow}}$ being the time-reversal symmetric~\footnote{These toy cat states resemble the prescription used in Ref.~\cite{morningstar_avalanches_2022} to unmix many-body resonances.}.

\begin{figure}[t!]
    \centering
    \includegraphics[width=1\columnwidth]{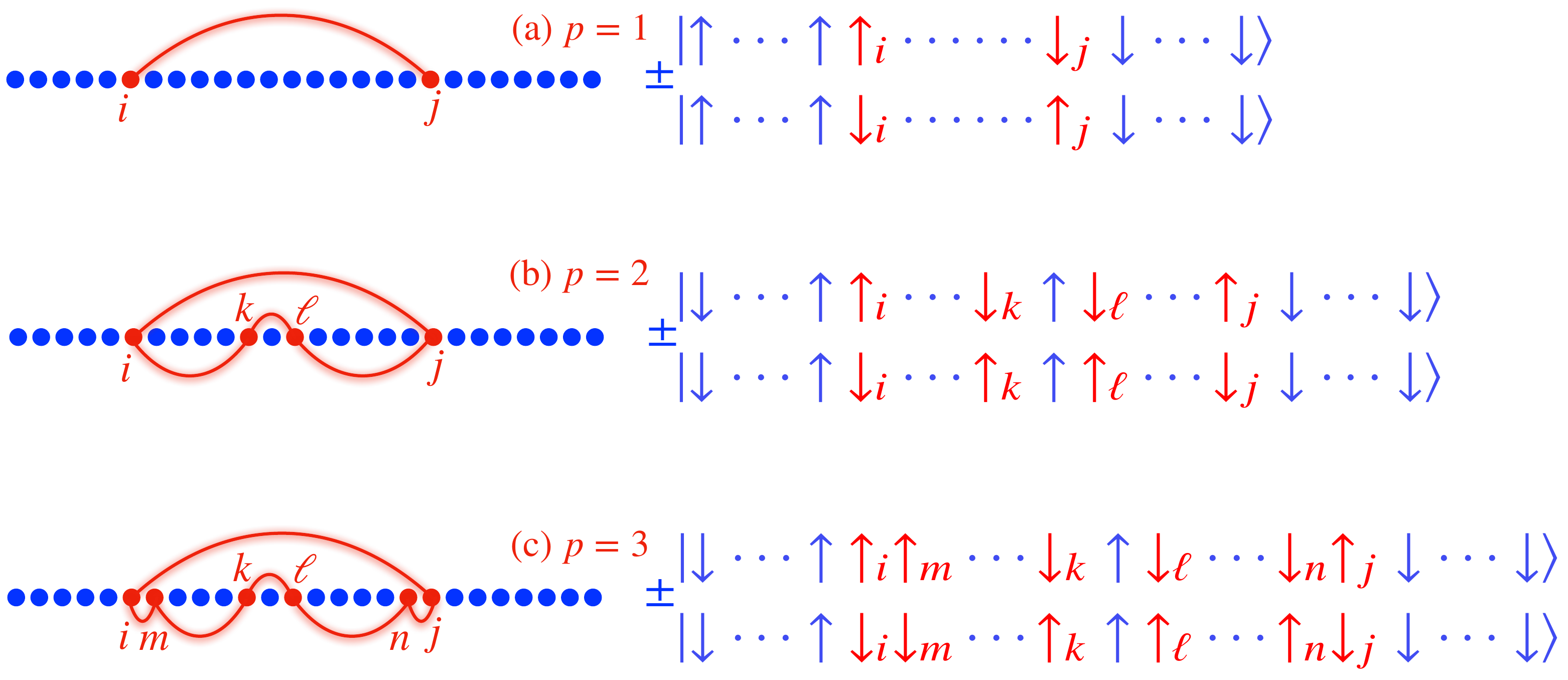}
    \caption{Illustration of some perfect cat states, described by the general form ${\ket{{\Phi_{2p}^{\pm}}}}$ of Eq.~\eqref{eq:Phi2p} with $\theta=\pi/2$. Panel (a) corresponds to the peculiar $p=1$ case, which has the  Bell-type of the $\Psi$ form Eq.~\eqref{eq:Psi}, while panels (b-c) show examples for $p=2,\,3$.}
    \label{fig:sketch_cats}
\end{figure}

Interestingly, the toy cat states of the form ${\ket{\Phi_{2p}^{\pm}}}$ given in Eq.~\eqref{eq:Phi2p} are rather general as they can also describe $|\Psi \rangle$ states in Eq.~\eqref{eq:Psi} by simply taking $p=1$. 
Fig.~\ref{fig:sketch_cats} illustrates three maximally entangled/correlated $(\theta=\pi/2)$ cases, for $p=1,\,2,\,3$. 
\subsubsection{Physical observables and measures of catness}
\label{sec:predictiontoystates}
\begin{figure}[b]
    \centering
    \includegraphics[width=\columnwidth]{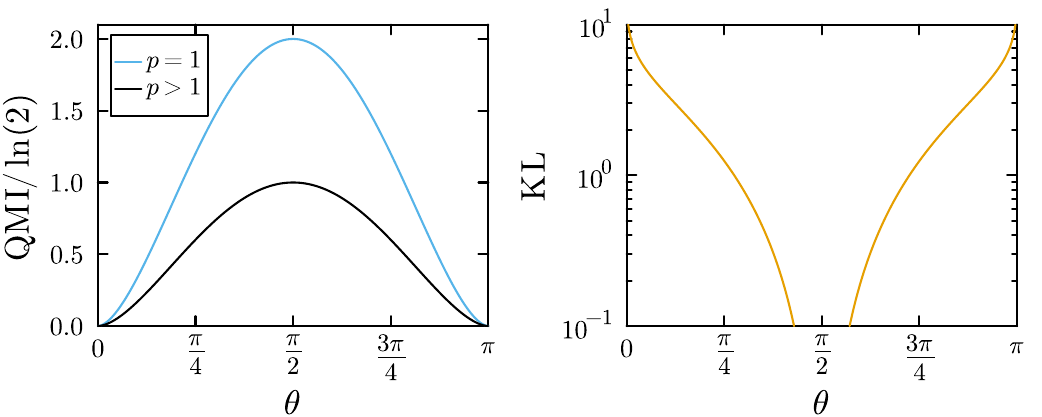}
    \caption{Quantum mutual information (left) Eq.~\eqref{eq:QMIcat}, and Kullback-leibler divergence (right) Eq.~\eqref{eq:KLCat}, plotted  as a function of the angle $\theta$ controlling the catness of the toy cat states. Increasingly good catness is signaled by the vanishing of KL for $\theta\to\pi/2$, and a QMI reaching its upper bounds, i.e. $\ln 4$ for two-spin resonant states ($p=1$), or $\ln 2$ for more $p>1$.}
    \label{fig:KLQMICats}
\end{figure}

\noindent{\it{(i) One- and two-point observables---}} 
The above results, displayed in Tab.~\ref{tab:catstates} for the simpler 2-site wave-functions Eqs.~\eqref{eq:Psi} and \eqref{eq:Phi} can be used for the more general ansatz of Eq.~\eqref{eq:Phi2p}. The case $p=1$ corresponds exactly to ${\ket{\Psi}}$, and $p\ge 2$ to ${\ket{\Phi}}$ but with a negative sign for $C^{zz}_{ij}$ if the concerned sites $(ij)$ have anti-aligned spins in the cat state expression, like for instance sites $i$ and $n$ in Fig.~\ref{fig:sketch_cats}~(c). One should emphasize again that the important distinction between different cat forms is that the transverse correlations $C^{xx}_{ij}$ are non zero {\it{only}} if $p=1$.\\
\\
{\it{(ii) Quantum mutual information---}} The two-site quantum mutual information (QMI)  also allows to discriminate between the different types of cat pairs. Indeed, 
QMI$_\theta$ is given by (see Appendix~\ref{app:qmi})
\be
{\mathrm{QMI}}_\theta=Q_{p}\left[\ln \left(\frac{2}{\sin\theta}\right) -\frac{\cos\theta}{2}\ln\left(\frac{1+\cos\theta}{1-\cos\theta}\right)\right],
\label{eq:QMIcat}
\ee
with the prefactors $Q_1=2$ and $Q_p=1$  for $p>1$, as seen in Fig.~\ref{fig:KLQMICats}. One immediately notices that QMI is upper bounded by ${\rm{QMI}}_{\pi/2}=\ln 2$ for $p>1$ and by ${\rm{QMI}}_{\pi/2}=\ln 4$ for $p=1$.\\
\\
{\it{(iii) First catness measure: the Kullback-Leibler divergence---}}
In a given computational basis $\{{\ket{\alpha_j}}\}$, the Kullback-Leibler (KL) divergence~\cite{kullback_information_1951} between two normalized states ${\ket{\phi_{1,2}}}$, expressed in this basis by 
\be
{\ket{\phi_m}}=\sum_{j=1}^{{\cal{N}}_{\rm H}} \sqrt{p^{(m)}_j}\exp\left(i\varphi_{j}^{(m)}\right){\ket{\alpha_j}},
\ee
is given by the following expression
\be 
{\rm{KL}}^{[12]}=\sum_{j=1}^{{\cal{N}}_{\rm H}} p^{(1)}_j\ln\left(\frac{p^{(1)}_j}{p^{(2)}_j}\right).
\label{eq:KL_Def}
\ee
Throughout this paper, we consider only the case where $\{{\ket{\alpha}}\}$ is the standard $\{S^z\}$ basis (i.e. $| \!\! \uparrow \downarrow \downarrow \dots \rangle$).\\

While any two eigenstates are orthogonal by definition, if they form a strongly entangled cat pair ($\theta\sim \pi/2$), the probabilities $p_j^{(1,2)}$ can be as close as possible, so their KL divergence, which quantifies how close two quantum states are, can be very small. Therefore, the KL is expected to provide a very good estimate of the catness of a pair. For our toy cat state ansatz ${\ket{{\Phi_{2p}^{\pm}}}}$, Eq.~\eqref{eq:Phi2p}, parametrized by the angle $\theta$, it is straightforward to compute the relative KL divergences within the pair, which is given for any value of $p\ge 1$, by
\be
{\rm KL}^{[\pm]}_{\theta}=-2\cos\left(\theta\right)\ln\left[\tan\left(\frac{\theta}{2}\right)\right].
\label{eq:KLCat}
\ee
This is a very sensitive witness of catness, rapidly vanishing when  $\theta \to \pi/2$ (see Fig.\ref{fig:KLQMICats}). More precisely, if $\theta=\pi/2\pm \delta$, for small $\delta$ we get ${\rm KL}^{[\pm]}\approx 2\delta^2$.\\
\\
{\it{(iv) Second catness measure: the $\Sigma$ test---}} 
Another  important quantity that will be used below to quantify the ``catness'' of an eigenstate pair is the off-diagonal matrix element
\be
\Sigma_{i}^{mn}={\bra{\phi_m}} \sigma_i^z{\ket{\phi_n}},
\ee
which we define as the "$\Sigma$ test" for a pair of eigenstates ${\ket {\phi_m}}$ and ${\ket{\phi_n}}$, where $i$ is a fluctuating site.
For standard MBL eigenstates that are adiabatically connected (via quasi-local unitaries) to a simple product state, such  off-diagonal elements are vanishingly small, while for resonant cat pairs of the above form Eq.~\eqref{eq:Phi2p}, we expect large ${\cal{O}}(1)$ values. A simple calculation gives
\be
\left|\Sigma_{i}^{\pm}\right|=\sin\theta,
\ee
the sign of $\Sigma_{i}^{\pm}$ being determined by the orientation of the local magnetization $\langle S_i^z\rangle / |\langle S_i^z\rangle|$. Remarkably, for perfect cats ($\theta=\pi/2$), we have $\left|\Sigma^{\pm}_i\right|=1$.

\subsubsection{Two illustrating examples}
\label{sec:examples}
We illustrate the ``catness'' features by looking at yet another set of examples in Table~\ref{tab:samples}, which clearly  exhibit all the expected characteristic of cat state pairs \footnote{Sample 2 is discussed in Appendix~\ref{app:examplemore}}. We recall that the ``selection'' (which is more explicitly detailed below in Sec.~\ref{sec:selectionalgorithm}) is firstly built on the criterion of having strong $|C_{L/2}^{zz}|$.

\begin{table}[h!]
    \centering
    \begin{tabular}{l|c|c||c|c|}
& \multicolumn{2}{c||}{{\rm{Sample}} 3}& \multicolumn{2}{c|}{{\rm{Sample}} 4} \\
\hline
$\epsilon$& 0.584447137 &0.584447089 & 0.473941 &0.473937\\
\hline  
        $\Delta\epsilon$ & \multicolumn{2}{c||}{$4.78\times 10^{-8}$} & \multicolumn{2}{c|}{$4.23\times 10^{-6}$}\\
         \hline
         $4C_{L/2}^{zz}$& -0.98427 &-0.98409 & -0.95596 &-0.96008\\
         \hline
         $4C_{L/2}^{xx}$& $-1.7\times 10^{-6}$ &$-2\times 10^{-6}$ & 0.97774 & -0.97984 \\
         \hline
         $2\langle S_i^z \rangle$& 0.07001 & -0.07163  & -0.14827   & 0.15968 \\
         $2\langle S_{i+L/2}^z \rangle$& -0.06717  & 0.07329 & 0.16136   & -0.15392 \\
                  \hline
         QMI &0.6568 &{0.6574} &1.3011 & {1.3117} \\
                  \hline
         KL & \multicolumn{2}{c||}{0.01014} & \multicolumn{2}{c|}{0.0531}\\
         \hline
         $\Sigma_i^\pm$ & \multicolumn{2}{c||}{0.9956} & \multicolumn{2}{c|}{0.9739}\\
         $\Sigma_{i+L/2}^\pm$ & \multicolumn{2}{c||}{-0.9885} & \multicolumn{2}{c|}{-0.9837}\\
         \hline
         $4C^{zz}_{L/2}-\Sigma_i^\pm\Sigma_j^\pm$& -0.00012 & 0.00006 & 0.002 & -0.002\\
         \hline
         \hline
    \end{tabular}
    \caption{Two examples of nearly degenerate cat-state pairs found in the middle of the many-body spectra of two representative samples at strong disorder ($h=20$). $L=16$ for Sample 3 (left), and $L=14$ for Sample 4 (right).}
    \label{tab:samples}
\end{table}

Both samples have very large $|C^{zz}_{L/2}|\sim 1$ as it is the main selection criterion, but interestingly the transverse correlations are completely different between the two samples: close to zero for Sample 3, they are maximal and {\it{staggered}} $\approx \pm 1$ for Sample 4. Comparing with Table~\ref{tab:catstates} we see that Sample 3 seems to be of type $\ket{\Phi^{\pm}}$, Eq.~\eqref{eq:Phi2p},  while Sample 4 is of type $\ket{\Psi^{\pm}}$. Or, using the more general ansatz of Eq.~\eqref{eq:Phi2p}, Sample 3 has $p>1$, while $p=1$ for Sample 4. This is perfectly confirmed by the QMI which is close to $\ln 2$ for Sample 3 and $\ln 4$ for Sample 4. 

In both cases, the local magnetizations are small and opposite on the fluctuating sites, the KL divergences within the cat pairs are very small, and finally the $\Sigma$s are very close to $\pm 1$, also signaling the antiferromagnetic pattern.\\

In addition, we find very small gaps between the cat states, typically on the smaller side when compared to the natural many-body level spacing at $h = 20$ for these sizes (see Sec.~\ref{sec:CatSpectroscopy}).

\subsection{Catness of states with long-range correlations}
\label{sec:Catness}
Beyond the examples of Sec.~\ref{sec:LRCCSToy}, we now ask if the states with large, long-range correlations are indeed well described by the proposed framework of toy cat states, and contrast the results with more generic (MBL) states in the many-body spectrum, through the lens of these observables.
\subsubsection{Selection algorithm}
\label{sec:selectionalgorithm}

For each disordered sample, we diagonalize the system and obtain the full set of eigenstates $\{{\ket{\phi_n}}\}$ ($n=1,\ldots,{\cal{N}}_{\rm H}$). Focusing on strong mid-chain $zz$ correlations at distance $L/2$, we first search for the largest $|C_{L/2}^{zz}|$ over the entire many-body spectrum, and select the corresponding eigenstate ${\ket {\phi_p}}$ if it satisfies the condition
\be
|C_{L/2}^{zz}|\ge C_{\star}.
\label{eq:C1}
\ee

This strong $zz$ correlation occurs between two sites $(i,j)$, where $|j-i|=L/2$. We then search in the spectrum for potential states ${\ket{\phi_q}}$ ($q\neq p$) such that Eq.~\eqref{eq:C1} is also satisfied for the same couple $(i,j)$. In the very rare cases where we find more than one such eigenstate, in order to capture the most probable cat partner for ${\ket {\phi_p}}$ (if it exists), we then compute the relative KL divergences between ${\ket {\phi_q}}$ and  ${\ket {\phi_p}}$, and only keep the candidate state that gives the minimum KL.

At present, we do not impose any further conditions on either the KL, the transverse correlations or $\Sigma$. Our analysis will show that they highlight signatures of the cat states, and they will be used in Sec.~\ref{sec:CatAnatomy}.

\subsubsection{KL divergences within potential cat pairs}
\noindent{\it{(i) KL without state selection--- }} For comparison, we briefly recall the main properties of the KL divergence between \emph{neighboring} eigenstates (denoted ${\rm KL}_{\rm NN}$ below) for the \emph{whole} spectrum, as the disorder strength $h$ is varied. This object provides a very useful tool for estimating the level of correlations between different eigenstates~\cite{luitz_many-body_2015,pino_ergodic_2019,khaymovich_fragile_2020,bahovadinov_many-body_2022}. While at weak disorder, the average $\overline{{\rm KL}_{\rm NN}} = 2$ \cite{pino_from_2019}, at stronger $h$ (in the MBL phase) it becomes extensive $\overline{{\rm KL}_{\rm NN}} \propto L$~\cite{luitz_many-body_2015}. 

\begin{figure}[t]
    \centering
    \includegraphics[width=0.65\columnwidth]{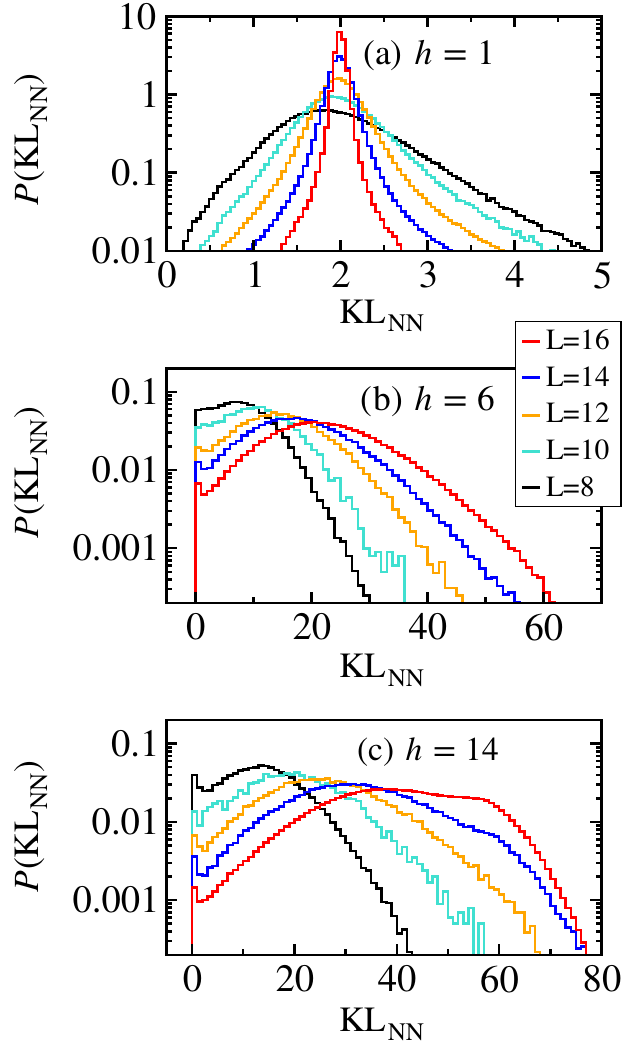}
    \caption{NN KL histograms over the entire spectrum, for small chains $(L=8,\,10,\,\cdots\,,\,16)$. At stronger disorder, a clear peak at very small KL appears. Note the numerical precision starts to be reached for some samples showing very large values of KL, typically above $\sim 60$.}
    \label{fig:KLFull}
\end{figure}

Fig.~\ref{fig:KLFull} shows the full distributions corresponding to Eq.~\eqref{eq:KL_Def} for the $\{S^z\}$ basis at three disorder strength. In panel (a) at $h=1$ the distributions indeed become narrower, peaking around ${\rm KL}_{\rm NN} = 2$ as the size increases. In contrast, at stronger disorder (b-c) $h = 6,14$, the distributions are much broader and their variance increases with the system size. Nevertheless we find that the quantity is self-averaging, as $P({\rm KL}_{\rm NN}/L)$ has a constant variance (data not shown).

The most interesting feature for the following is the intriguing presence at strong disorder of a significant upturn of the histograms for KL $\to 0$, visible in Fig.~\ref{fig:KLFull} (b-c). 
This indicates that some adjacent eigenstates present strong and pronounced similarity, in contrast to standard MBL states which are uncorrelated and exhibit a volume-law KL.  This feature at KL $\to 0$ is, as we argue below, a signature for the presence of cat state pairs in the many-body spectrum. 
Note that cat states need not be nearest neighbors in the spectrum (as already seen for some cat states in Fig.~\ref{fig:spectrum}) -- we will discuss their spectral distance in Sec.~\ref{sec:CatAnatomy}, but the majority of them are nearest-neighbors, up to the point that their presence becomes noticeable in the distribution of $P({\rm KL}_{\rm NN})$. The conclusion that emerges at this stage is that a special class of very similar nearest-neighbour eigenstates is definitely present and stable throughout the strong disorder regime of the phase diagram, regardless of system size $L$ and disorder $h$.\\
\\
{\it{(ii) Similarity analysis of candidate cat-states--- }} We now turn to a detailed analysis of the presumed cat-like states, as obtained from our filtering algorithm described in Sec.~\ref{sec:selectionalgorithm}. By treating a large number of disorder samples, we have of the order of a few tens of thousands of candidate cat-states for the smallest chains, and a few thousands for the largest systems. 
 \begin{figure}[t!]
    \centering
    \includegraphics[width=.85\columnwidth]{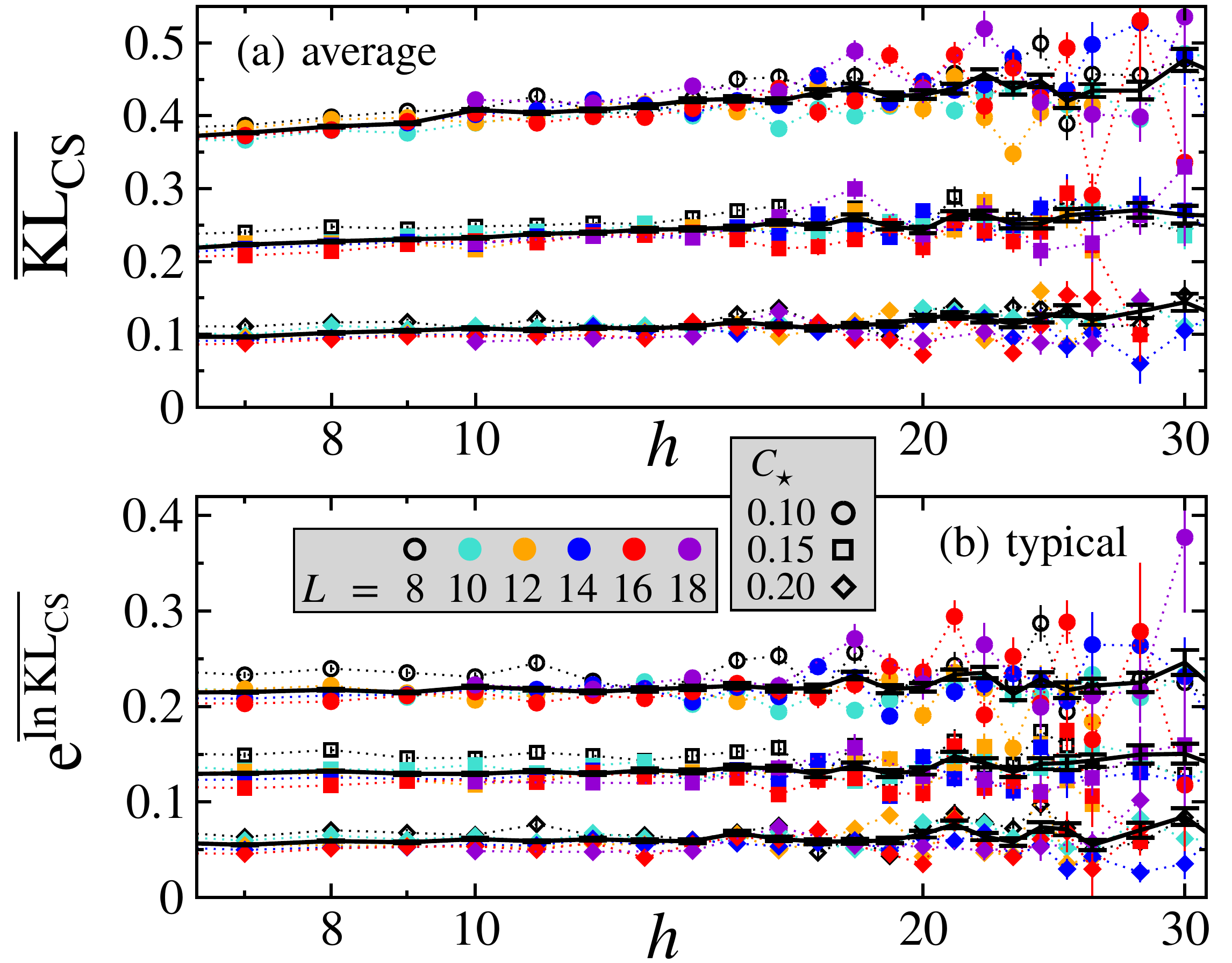}
    \caption{Average (top) and typical (bottom) values of the ${\rm KL}$ for potential cat states (Sec. \ref{sec:selectionalgorithm}), for all available system sizes and fields, and for three possible values of the threshold $C^{zz}_{L/2} \geq C_{\star}$.}
    \label{fig:KLCats_vs_h}
\end{figure}

\paragraph{First catness check.} We first analyze the KL$_{\rm CS}$ of (presumed) cat states in Fig.~\ref{fig:KLCats_vs_h}, where we show their $h$ dependence, for both average and typical divergences, for all available system sizes, restricting our analysis to strong disorder $h>6$.
Clearly, this class of eigenstates shows a very different behavior from the nominal $h$-dependence of KL$_{\rm NN}$: while KL$_{\rm NN}\gg 1$ grows with both $L$ and $h$, we find that KL$_{\rm CS}\ll 1$, and that it shows almost no dependence on either $h$ or $L$. This indicates that pairs of eigenstates with very large $C^{zz}_{L/2}$ are extremely similar, and also further strengthens our intuition that cat states are the eigenstates contributing to the small-KL peak in the distribution $P($KL$_{\rm NN})$ in Fig.~\ref{fig:KLFull}.

Another significant difference is the fact that average and typical values of KL$_{\rm CS}$ consistently differ (by a factor of about $2$), in contrast to KL$_{\rm NN}$ for which they coincide.
Finally, an additional conclusion of Fig.~\ref{fig:KLCats_vs_h} is that the main dependence of KL$_{\rm CS}$ is on the threshold $C_\star$, which is set beforehand in the filtering process.

 \begin{figure}[b!]
    \centering
    \includegraphics[width=.8\columnwidth]{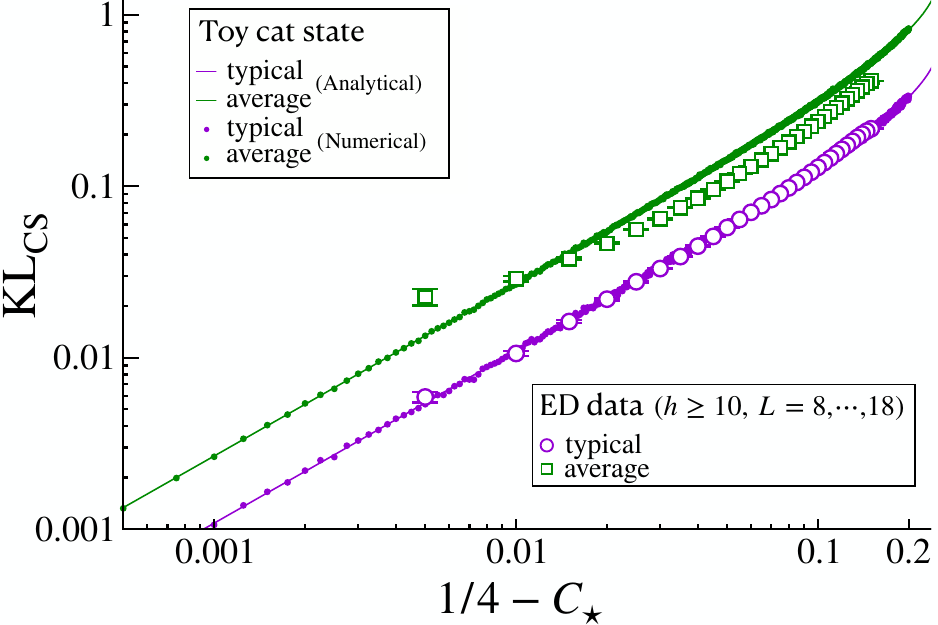}
    \caption{KL of the cat states plotted against the threshold correlation $C_\star$. Analytical expressions Eq.~\eqref{eq:AVG_and_TYPKL} (full lines) are compared with ED data (symbols), collected over all lengths $L=8,\cdots,18$ and over a large part of the rare-event regime $h\ge 10$. Small filled symbols show numerical evaluation of the toy cat state integrals Eq.~\eqref{eq:avgKLcs} and Eq.~\eqref{eq:typKLcs}.}
    \label{fig:KLCats}
\end{figure}

\paragraph{Comparison to predictions for toy cat states---} 

We now try to account for these observations based on the toy cat states introduced in Sec.~\ref{sec:toystates}. 
We start with the dependence of KL$_{\rm CS}$ on the value of the threshold $C_\star$ in Fig.~\ref{fig:KLCats} by representing, for comparison convenience, the typical and average values of KL$_{\rm CS}$ against $1/4-C_\star$ (i.e. the deviation from perfect catness $C_{\rm max}=1/4$). In the limit $C_\star \to 1/4$, close to maximally entangled cat pairs (i.e. with an angle $\theta\approx \pi/2$), we expect KL$_{\rm CS}$ to vanish, after expanding Eq.~\eqref{eq:KLCat} close to $\pi/2$, as
\begin{equation}
    {\rm KL}_{\rm CS}^{[\pm]}(\theta)\approx 2\delta^2-\delta^4,\quad(\delta=\frac{\pi}{2}-\theta),
    \label{eq:KL_expansion}
\end{equation}
where $\delta$ measures the distance form perfect catness. In order to make a quantitative comparison of our ED data in Fig.~\ref{fig:KLCats} with simple analytical predictions such as Eq.~\eqref{eq:KL_expansion}, we need to model the statistical averaging.

The selection of correlations within the interval $[C_\star ; 1/4]$ correspond in the toy cat states modelization to an angle $\theta_\star \le \theta< \pi/2$ (assuming $\theta\le \pi/2$ without loss of generality).  The simplest choice is to consider a uniform distribution for $\theta \in [\theta_\star ; \pi/2]$, where the angle $\theta_\star$ is such that  \be 
C_\star=\frac{1}{4}\sin^2\left(\theta_\star\right).
\ee
This allows to obtain both average and typical ${\rm KL}_{\rm CS}^{[\pm]}(\theta_\star)$ (see Appendix \ref{app:KLcat} for details) which take the forms:
\be
\exp\left({\overline{\ln {{\rm{KL}}}_{\rm CS}}}\right)\approx \frac{2}{e^2}\left[\frac{\pi}{2}-\theta_\star\right]^2\approx \frac{3}{e^2}\overline{{\rm KL}_{\rm CS}^{[\pm]}(\theta_\star)}.
\label{eq:AVG_and_TYPKL}
\ee

Fig.~\ref{fig:KLCats} shows that the dependence of the typical ${\rm KL}_{\rm CS}$ on the threshold is very well captured by the toy cat states. Although not perfectly, the behavior of the average KL is qualitatively also well captured by the toy states, except at very large correlations where we have very little statistics. Globally, the toy cat states model reproduces very consistently the observed similarity between pairs of strongly-correlated eigenstates. We further note the strong correlation between low-values of ${\rm KL}$ and good (toy state) catness, which will be used sometimes below as an additional cat states filtering criterion.

\subsubsection{The $\Sigma$-test}
\noindent{\it{(i) Distribution of the $\Sigma_i^\pm$---}}~For a candidate pair of eigenstates $\ket{\Phi^{\pm}}$ , the computation of  
\be
\Sigma_{i}^{\pm}=\bra{{\Phi^+}}\sigma_i^z\ket{{\Phi^-}}
\label{eq:Sigma}
\ee
is a very sensitive test of their catness. Indeed, for the sites $i$ that exhibit strong correlations, the toy description of cat states Eq.~\eqref{eq:Phi2p} gives $\Sigma_i^\pm=\Sigma_{i+L/2}^\pm=\sin\theta$ (instead of $\simeq 0$ for standard MBL states). Assuming again a simple uniform distribution for the angles $\theta\in [\theta_\star\,;\,\pi/2]$, one easily arrives at the following expression for its probability distribution
 \begin{figure}[b!]
    \centering
    \includegraphics[width=1\columnwidth]{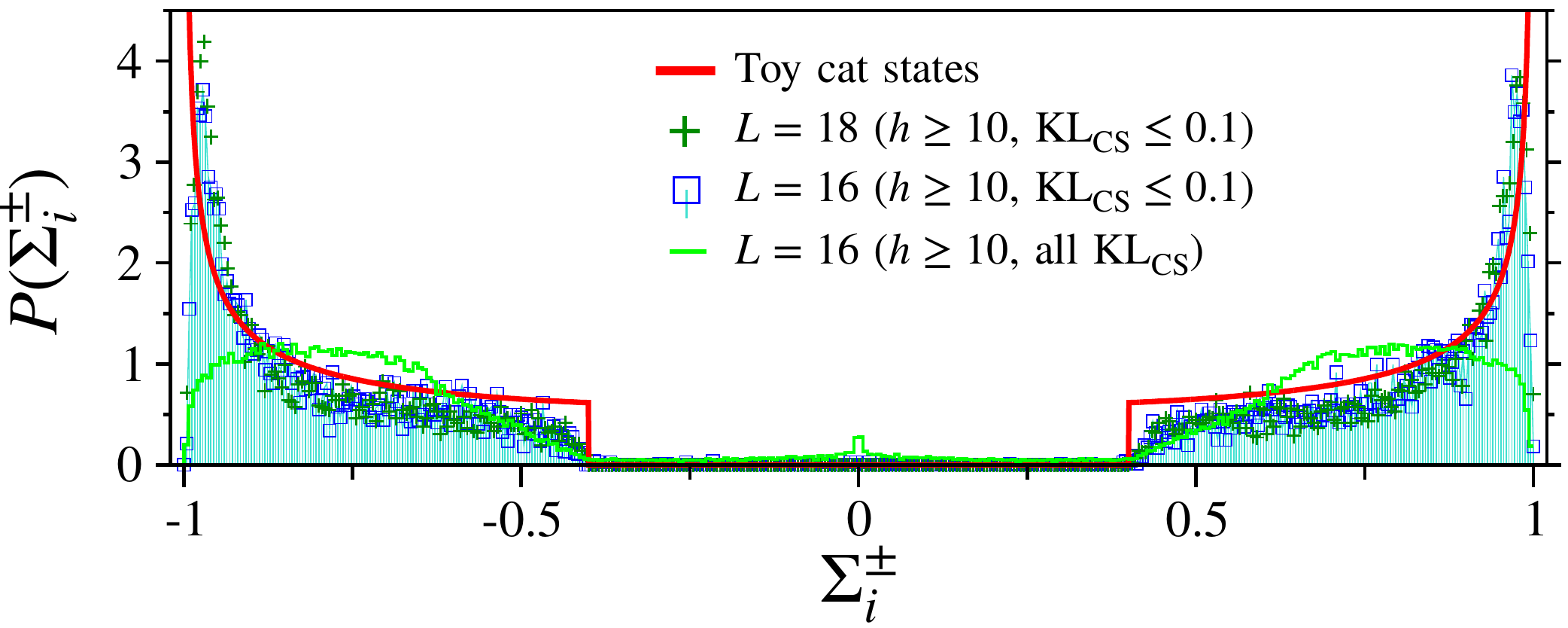}
    \caption{Distribution of the $\Sigma_i^\pm$, Eq.~\eqref{eq:Sigma}, shown for the two largest systems ($L=16,\,18$), and compared to the expression Eq.~\eqref{eq:TMSigma} for toy cat states. Numerical data, collected for $h\in [10,\,30]$ over all available cat pairs ($\approx 20000$ for $L=16$ and $\approx 7000$ for $L=18$ for $C_\star=0.1$) having a small Kullback-Leibler divergence (${\rm{KL_{CS}}}<0.1$) show a remarkable agreement with the analytical prediction (red line). The agreement is less good if one keeps all values of KLs (green curve).}
    \label{fig:Psigmas}
\end{figure}
\begin{equation}
P(\Sigma^{\pm})=\frac{2\Theta\left(|\Sigma^{\pm}|-4C_\star\right)}{(\frac{\pi}{2}-\theta_\star)\sqrt{1-\left(\Sigma^{\pm}\right)^2}},
    \label{eq:TMSigma}
\end{equation}
which is shown in Fig.~\ref{fig:Psigmas} together with ED results. The comparison is very interesting: the toy cat states prediction describes quite well the ED data when considering the pairs with low KL$_{\rm CS}\le 0.1$. When all KLs are kept, the shape of the distribution gets more rounded, with an additional small peak at zero, characterized  more precisely below.\\
\\
{\it{(ii) Correlation between $C_{ij}^{zz}$ and $\Sigma_i^\pm \Sigma_j^\pm$---}}
It is easy to relate the connected longitudinal correlation function between sites ($i,j$) in a given eigenstate ${\ket {\phi_m}}$ to a sum of product of $\Sigma_i$:
\begin{eqnarray}
    4C_{ij}^{zz}\Bigr|_{m}&=&\bra{\phi_m}\sigma_i^z\sigma_j^z\ket{\phi_m} -\bra{\phi_m}\sigma_i^z\ket{\phi_m}\bra{\phi_m}\sigma_j^z\ket{\phi_m}\nonumber\\
    &=&\sum_{n\neq m}\bra{\phi_m}\sigma_i^z\ket{\phi_n}\bra{\phi_n}\sigma_j^z\ket{\phi_m}\nonumber\\
    &=&\sum_{n\neq m}\Sigma_{i}^{nm}\Sigma_{j}^{nm}.\label{eq:Cphim}
\end{eqnarray}
For typical (short-range correlated) MBL eigenstates, we expect that the sum in Eq.~\eqref{eq:Cphim} will remain exponentially suppressed with $|i-j|$. In contrast, for a cat state ${\ket{\phi_m}}$ having a large correlation, the sum will be dominated by the two protagonists of the pair, say ${\ket{\phi^{\pm}}}$, providing an estimate 
\be
4C_{ij}^{zz}\Bigr|_{\phi_\pm}\approx \Sigma_{i}^{\pm}\Sigma_{j}^{\pm}={\cal{O}}(1).
\label{eq:CzzSigmas}
\ee
For the ansatz pair states Eq.~\eqref{eq:Phi2p} parametrized by $\theta$, Eq.~\eqref{eq:CzzSigmas} is an exact equality with the ${\cal{O}}(1)$ constant equal to $\sin^2\theta$. 

 \begin{figure}[t!]
    \centering
    \includegraphics[width=1\columnwidth]{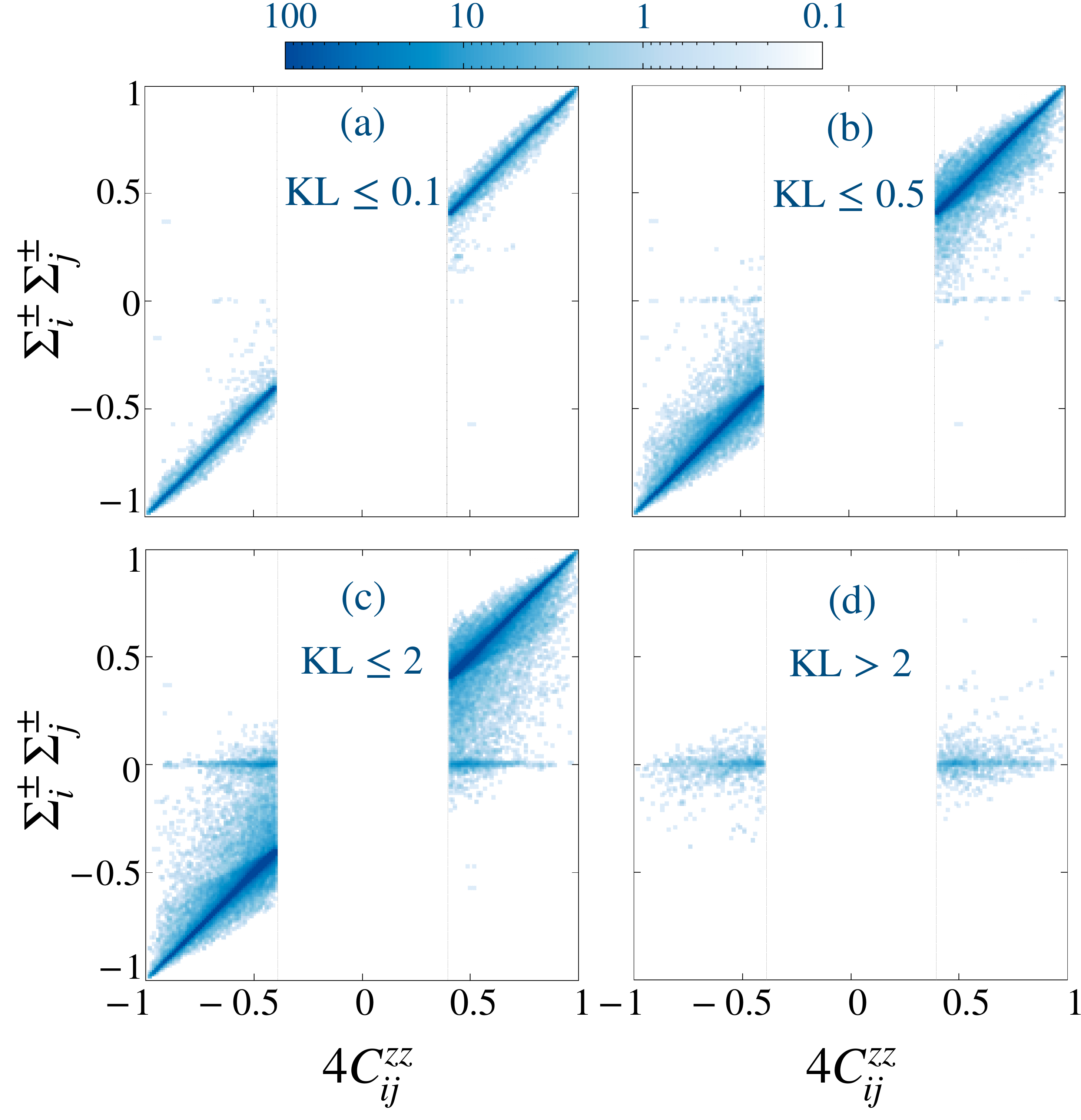}
    \caption{Cross correlations between  $C_{ij}^{zz}$ and $\Sigma_{i}^{\pm}\Sigma_{j}^{\pm}$ for the mid-chain correlations $(|i-j|=L/2)$ of candidate cat states, filtered using $C_\star=0.1$ (dotted vertical lines at $4|C_{ij}^{zz}|=0.4$). $L=16$ data are collected over $10^5$ random samples for disorder strength $h\in [10\,;\,30]$. 
    The 4 different panels show KL-filtered results, yielding various number of targeted eigenpairs $N_p$:
    (a) ${\rm{KL}}\le 0.1$ $N_p\approx 13000$, (b) ${\rm{KL}}\le 0.5$ $N_p\approx 35000$, (c) ${\rm{KL}}\le 2$ $N_p\approx 50000$, (d) ${\rm{KL}}>2$ $N_p\approx  1300$. For ${\rm{KL}}\le 2$  most of the points align along the bisector, but as KL increases, spurious events ($|\Sigma_{i}^{\pm}\Sigma_{j}^{\pm}|\ll 4|C_{ij}^{zz}|$) appear. No cat states are detected for ${\rm{KL}}>2$.}
    \label{fig:Correlations_Czz_Sigmas}
\end{figure}

For candidate cat states, testing Eq.~\eqref{eq:CzzSigmas} is another salient probe of catness, as shown in Fig.~\ref{fig:Correlations_Czz_Sigmas}, where we numerically demonstrate the very strong correlation between $C_{ij}^{zz}$ and the product $\Sigma_{i}^{\pm}\Sigma_{j}^{\pm}$, in line with the expected results for the toy cat states.
Interestingly, if one introduces an additional filtering with respect to the value of ${\rm KL}$, it clearly highlights its role in capturing ``genuine'' cat states. 
While for ${\rm KL}\leq 0.5$ the vast majority of the states align on a very clear correlation line, for KL$\leq 2$, we clearly obtain additional states which fall out of this line. A fortiori, the $\Sigma$-test basically fails for  KL$>2$, revealing an absence of genuine cat states for these large values of Kullback-Leibler divergences.

\section{Anatomy of the cat states}
\label{sec:CatAnatomy}

Thanks to the analysis in Sec.~\ref{sec:LRCCS}, we are able to identify and focus on cat states to study their spectral properties and microscopic structure. To capture genuine cat states, we only consider pairs with ${\rm KL}_{\rm CS} < 0.5$ from now on.  We find (data not shown) that most of the cat pairs ($70$ to $95\% $ depending on system size) have a spectral distance $d_s=1$, that is, are nearest neighbor energy states \footnote{The distribution of the spectral distances appears to be very singular, and could be reasonably well described by a fats power law, $P(d_s)\sim d_s^{-\gamma}$ with a rather large exponent $\gamma$, that we typically find in the range of $\sim 3-6$.}.
In the following, we discuss cat pairs with spectral distance $d_s=1$, thus focusing on the most resonant quasi-degenerate states.\\

\subsection{Spectroscopy of the cat states}
\label{sec:CatSpectroscopy}
\subsubsection{Density of (cat) states}

 \begin{figure}[b]
    \centering
    \includegraphics[width=.8\columnwidth]{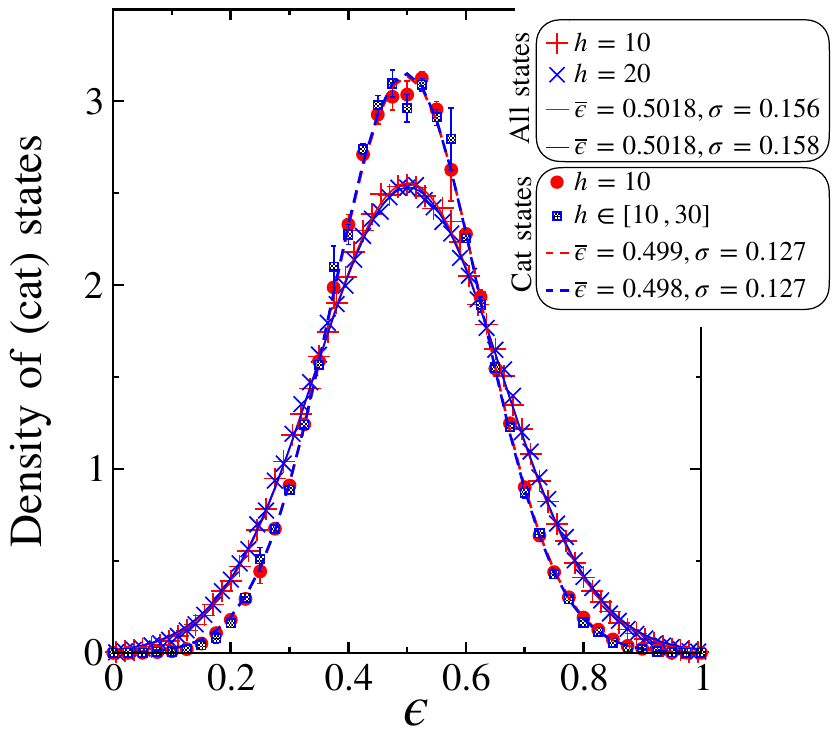}
    \caption{{Density of states for all eigenstates (cross symbols) or only cat states (full symbols) for a $L=16$ chain and two disorder strengths $h=10$ (red symbols) and $h=20$ (blue symbols).  Parameters of the Gaussian fits Eq.~\eqref{eq:DOCS} are given in the legend.}}
    \label{fig:DOCS}
\end{figure}
We first discuss the normalized density of states (DOS) and the density of cat states (DOCS), that we plot in Fig.~\ref{fig:DOCS}. They both exhibit Gaussian shapes, centered around $\epsilon=0.5$, with no quantitative dependence on $h$, at least in the regime of interest $h\ge 10$. However, there is a noticeable interesting distinction between the two densities as we observe that the DOCS is quantitatively more peaked at $\epsilon=0.5$, as compared to the DOS. Gaussian fits of the form 
\be
{\rm{DO(C)S}}(\epsilon)=\frac{1}{\sqrt{2\pi\sigma^2}}\exp\left(-\frac{(\epsilon-{\overline{\epsilon}})^2}{2\sigma^2}\right),
\label{eq:DOCS}
\ee
provide a very good description of the data, with a variance that is $\sim 40\%$ larger in the DOS than in the DOCS. While relatively modest, this narrowing is clearly visible, pointing the fact that the probability to find cat states is enhanced in the middle of the spectrum. This is also a sign that the level spacing of the cat pairs might be smaller than the natural spacing in the middle of the many body spectrum, a fact that we discuss more quantitatively now.

\subsubsection{Energy gaps and their distributions}
\label{sec:gaps}
\begin{figure}[t!]
    \centering
    \includegraphics[width=.9\columnwidth]{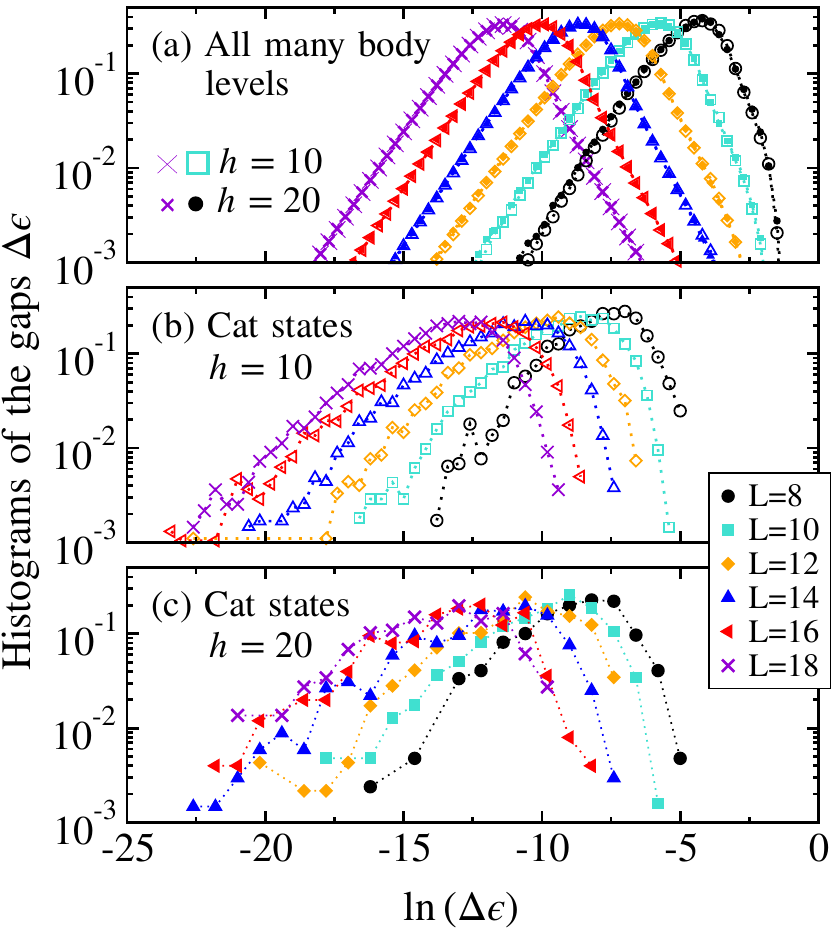}
    \caption{Histograms $P(\ln\Delta\epsilon)$ of the log of the energy gaps: (a) for the whole spectra, and (b-c) for cat states only. ED data are shown for two representative disorder strengths $h=10,\,20$. Histograms for the cats are built, for $L=(8,\,10,\,12,\,14,\,16,\,18)$ respectively from $\sim (6\cdot 10^3,\, 1.4\cdot 10^4,\,1.4\cdot 10^4,\, 2.4\cdot 10^4,\,1.9\cdot 10^4,\,7\cdot 10^3$) events for $h=10$ (b), and from $\sim (10^3,\, 1.5\cdot 10^3,\,1.2\cdot 10^3,\, 1.7\cdot 10^3,\, 600,\, 200$) events for $h=20$ (c).}
    \label{fig:HistoCatGaps}
\end{figure}

\noindent{\it (i) Histograms---}
We now consider the energy gaps in Fig.~\ref{fig:HistoCatGaps} where we show the histograms of  (the logarithm of) the many-body  level spacings for two representative (strong) values of  disorder. Panel (a) provides data for the full spectra, to be compared with the cat state gaps shown in panels (b) and (c), presented on the same scale. We first notice that the cat gaps are more broadly distributed than the full spectrum data, while their distributions do not seem to broaden upon increasing $L$. We also see that the order of magnitude of the cat gaps are significantly smaller than the natural level spacings (see below).  As detailed in the caption of Fig.~\ref{fig:HistoCatGaps}, the statistics of cat states become very poor for $h=20$ as we are dealing with extremely rare events, and the occurrence of cat states drops significantly for increasing disorder.\\

\noindent{\it(ii) Typical and average decays---}
A more detailed analysis of the scaling of the typical and average gaps of the cat states is given in Appendix~\ref{app:spectro}. The main result, as already seen from the histogram, is that the cat gaps can be much smaller than the natural level spacing, which typically goes as $1/{\cal{N}}_{\rm H}\sim 2^{-L}$. We find that both typical and average cat-gaps decay also exponentially with $L$, according to
\be
\Delta{\epsilon}_{_{\rm CS}}= \Delta_{0}(h)\exp\left(-\frac{L}{\xi_{\rm sp}(h)}\right),
\label{eq:Deltaepsilon}
\ee
with an almost $h$-independent spectral length $\xi_{\rm sp}\sim 1.7$ (slightly larger than the natural spectral length $(\ln 2)^{-1}\sim~1.44$), but with a very small prefactor $\Delta_0(h)$ that rapidly decreases with the disorder strength, presumably like a power-law. 
While both the typical and average cat gaps show a slightly slower exponential decay than the natural spacing, this effect is largely offset by the much smaller amplitude of the cat gaps, at least for the available system sizes. However, we are unable to conclude whether at very large sizes the natural gap will eventually become smaller than the cat gap, or not.
 
\subsection{Microscopic structure of the cat states}
\label{sec:CatMicroscopy}
\subsubsection{Fluctuating spins within cat eigenpairs}

To get more insight on the microscopic local structure of the nearly degenerate resonant cat states, we define the simple local expectation value within each doublet ${\ket{\Phi^{\pm}}}$
\be
 {\mathsf{N}}_{0} =L-\sum_{i=1}^{L}\left|{\bra{\Phi^+}} S_i^z {\ket{\Phi^+}}\,+\,{\bra{\Phi^-}} S_i^z {\ket{\Phi^-}}\right|,
\label{eq:N0}
\ee
which is expected to give an estimate of the number of fluctuating spins within a cat pair. For instance, using the ansatz
${\ket{{\Phi_{2p}^{\pm}}}}$ defined in 
Eq.~\eqref{eq:Phi2p}, the above sum simply yields ${\mathsf{N}}_{0}=2p$, i.e. the number of fluctuating sites.

For non-cat states, we expect two simple behaviors for ${\mathsf{N}}_0$. For  localized-bits type MBL eigenstates, $P(\mathsf{N}_0)$ is expected to be peaked at even values with an overall gaussian enveloppe centered around 
${\mathsf{N}}_0= L/2$, if one chooses adjacent eigenstates for  ${\ket{\Phi^\pm}}$ in Eq.~\eqref{eq:N0} (this does not mean that there are $L/2$ fluctuating sites, see App.~\ref{sec:PN0MBL}). For ergodic states and using again ${\ket{\Phi^\pm}}$ in Eq.~\eqref{eq:N0} as adjacent eigenstates, we expect ${\mathsf{N}}_0\to L$ (here, all sites fluctuate).

\subsubsection{Cat states anatomy: Histograms of the fluctuating sites}

 \begin{figure}[b]
     \centering
    \includegraphics[width=\columnwidth]{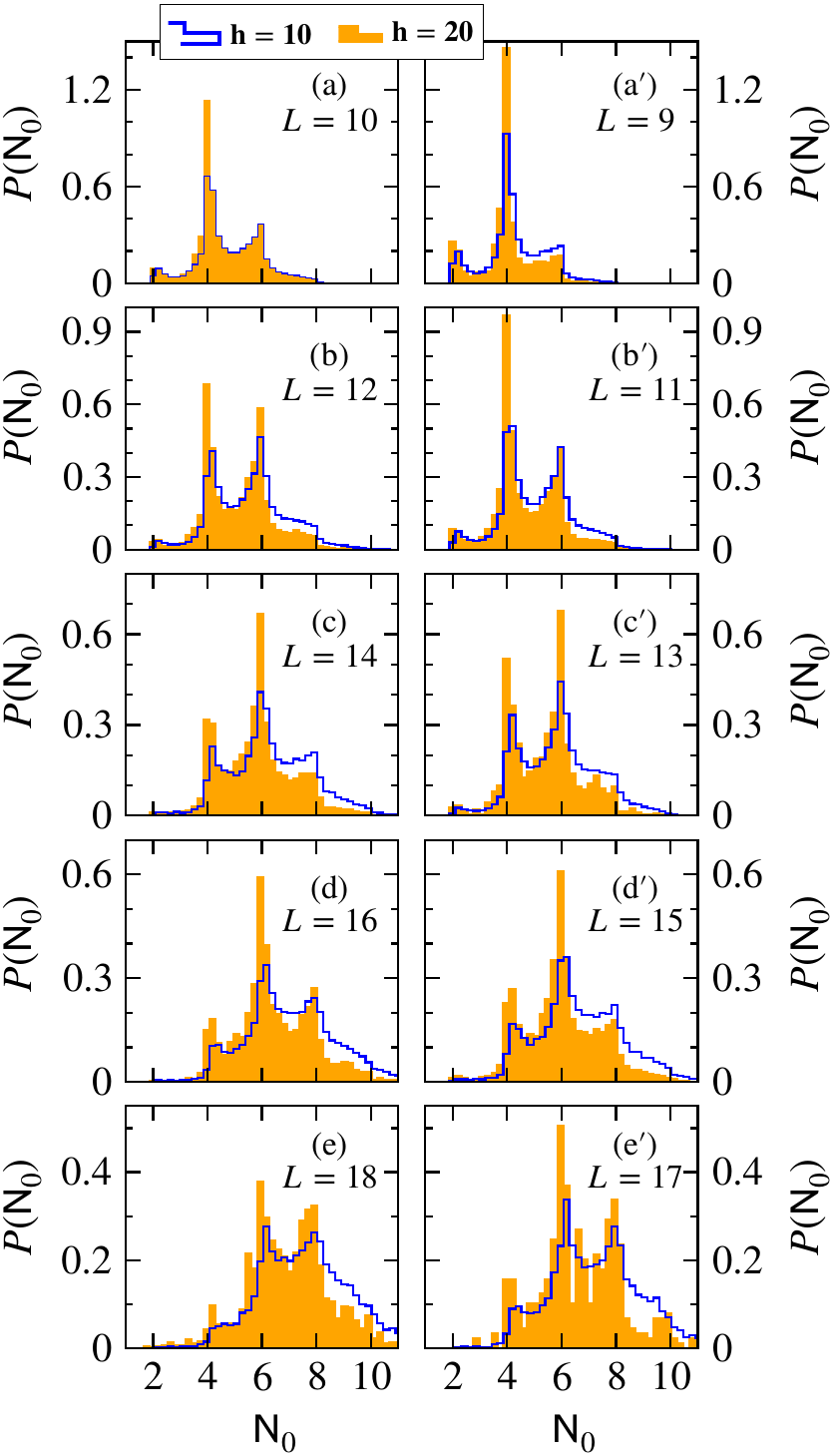}
    \caption{Histograms of the number of fluctuating sites $P({\mathsf{N}}_0)$, computed from Eq.~\eqref{eq:N0} for cat states only, shown for various system sizes, and two disorder strengths ($h=10,\,20$).}
    \label{fig:HistoN0}
 \end{figure}
Fig.~\ref{fig:HistoN0} shows the histograms of ${\mathsf{N}}_0$, computed over all the available cat pairs for $L=10,\,11,\,\cdots,\,18$. Data are shown for two representative disorder strengths $h=10$ and $h=20$.

We observe clear peaks at even numbers ($2p$) in the histograms $P({\mathsf{N}}_0)$ as the signature of the number of fluctuating sites involved in the resonant cat pairs. The peaks are deeper and more pronounced for $h=20$, as localized eigenstates get increasingly better described by ${\ket{\varphi_{L-2p}}}$ in the ansatz
${\ket{{\Phi_{2p}^{\pm}}}}$  of Eq.~\eqref{eq:Phi2p}. 

It is very interesting to notice that peaks gradually develop towards larger values of even integers when $L$ increases, meaning that an increasing number of fluctuating spins get involved in the cat states. This mechanism follows a non-trivial distribution of the number of pairwise fluctuations $P({\mathsf{N}}_0)$. Results for odd-numbered chains show similar effects, meaning that there is also an even number of fluctuating sites, and an odd number of frozen spins.

\subsubsection{Typical number of resonant spins}
\begin{figure}[h!]
     \centering
    \includegraphics[width=0.84\columnwidth]{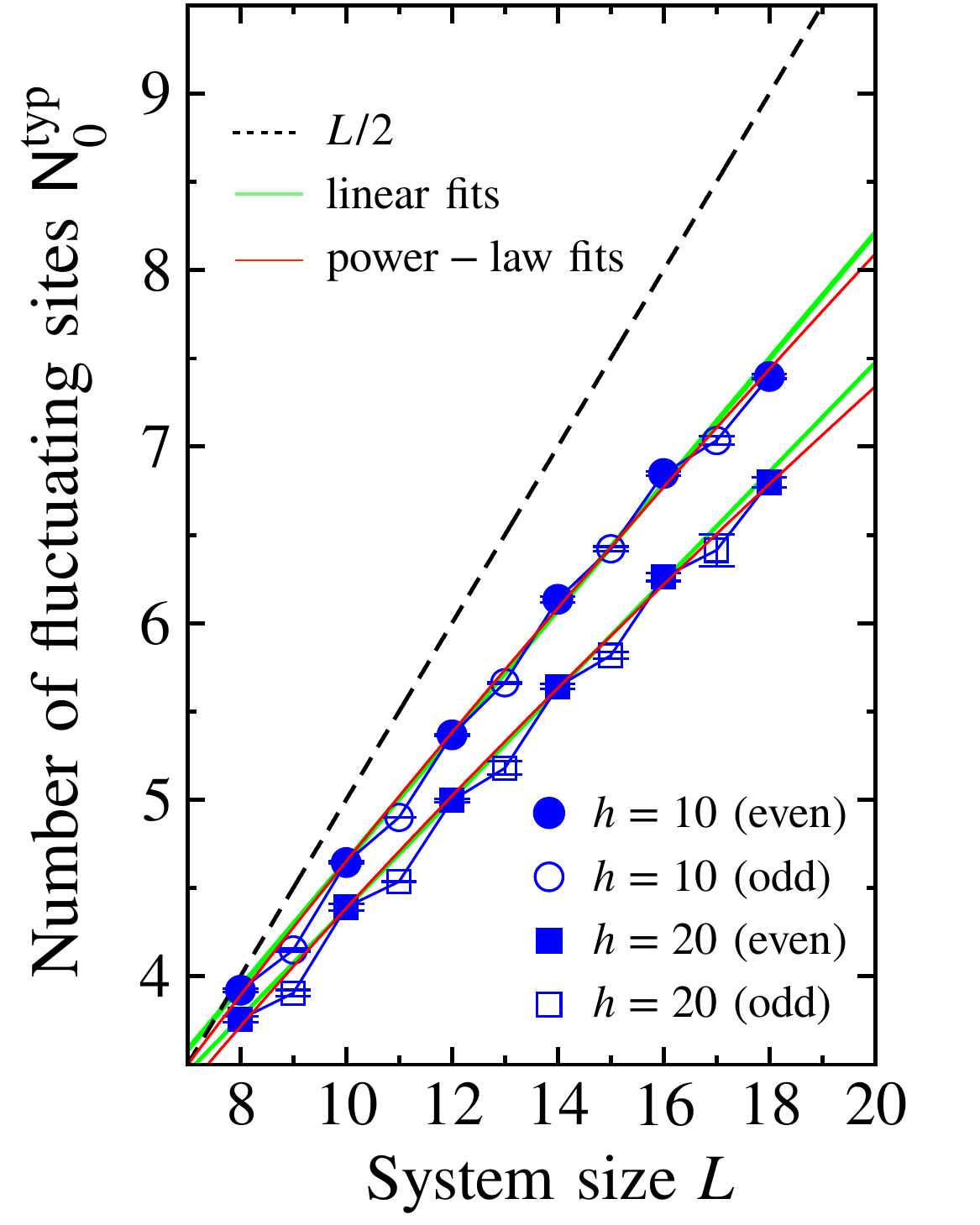}
    \caption{Typical number of fluctuating spins with the cat pairs, ${\mathsf{N}}_0^{\rm typ}$ computed from Eq.~\eqref{eq:N0}, plotted as a function of the system size $L$. Two values of disorder strengths ($h=10,\,20$) are shown for odd and even chains. Colored lines are linear (green) and power-law (red) fits, see Tab.~\ref{tab:fits} for the parameters. The dashed line at $L/2$ is shown for comparison.}
    \label{fig:N0}
 \end{figure}

Besides their distribution, it is also very instructive to  analyze the {\it{typical}} value of ${\mathsf{N}}_0$, providing an estimate for the typical number of sites that are involved in many-body resonances. The evolution of ${\mathsf{N}}_0^{\rm typ}$ with $L$ is shown in  Fig.~\ref{fig:N0}, for which we want to make a few remarks.

First, let us emphasize that the true number of fluctuating spins is probably better estimated at the strongest disorder, where the sum in Eq.~\eqref{eq:N0} may be more easily saturated by nearly perfectly polarized moments~\cite{khemani_obtaining_2016, lim_many-body_2016,dupont_many-body_2019, hopjan_many-body_2020,hopjan_detecting_2021,laflorencie_chain_2020, colbois_breaking_2023}. This is also evident from the histograms in Fig.~\ref{fig:HistoN0}, which have more weight at non-integer values for $h=10$ compared to $h=20$.

${\mathsf{N}}_0^{\rm typ}$ grows with $L$, but clearly slower than $L/2$, which would be expected from a naive perturbative coupling between spins at distance $L/2$. We have tried two types of fits (on even sizes only): a linear form ${\mathsf{N}}_0^{\rm typ}=x_{\rm fluc.}L+b$, and a power-law $\sim L^{\delta}$, the latter being slightly better, as perhaps visible from the small curvature in Fig.~\ref{fig:N0}. In Table~\ref{tab:fits} we provide the fitting parameters.
\begin{table}[h!]
\begin{center}
\begin{tabular}{c|c|c|}
&linear: $x_{\rm fluc}L+b$ & power-law: $\sim L^\delta$\\
\hline
$h=10$ & $x_{\rm fluc}=0.355(2),\,b=1.1(1)$ & $\delta=0.80(1)$\\
$h=20$ & $x_{\rm fluc}=0.309(3),\,b=1.3(1)$ & $\delta=0.74(1)$\\
\hline
\end{tabular}
\caption{\label{tab:fits}
Fitting parameters for even sizes data shown in Fig.~\ref{fig:N0} by green (linear) and red (power-law) lines.}
\end{center}
\end{table}

\subsubsection{Discussion}
The previous analysis provides us with valuable information about the local structure of this new type of long-range correlated eigenstates, understood in terms of resonant pairs of cat states. In fact, there is a very rich distribution of "possible paths" (in real space) that can be taken to establish long-range resonances in the system, as shown in Fig.~\ref{fig:HistoN0}, which gives an estimate of the number of intermediate spin flips that can occur in the cat pair. Interestingly, the most probable number of spin flips is significantly smaller than $L/2$, which tells us that there should be some long-range processes that can virtually couple distant spins across the chain. Recent results~\cite{Ha_many-body_2023} appear to be compatible with this conclusion, albeit in a different setup (Floquet system, with open boundary conditions and no magnetization conservation).

Typically, as reported in Fig.~\ref{fig:N0} and Tab.~\ref{tab:fits}, this number of spin flips is found to scale either linearly $\propto x_{\rm fluc}L$, with a fraction of fluctuating sites $x_{\rm typ}\sim 0.3$, or to follow a sublinear power law $\sim L^\delta$, with an exponent $\delta\sim 0.75$. This last possibility could be tentatively understood as the result of a fluctuating backbone with a fractal dimension $\delta<1$~\cite{potter_universal_2015}, an interpretation that may be consistent with proposed scenarios of rare fractal thermal inclusions within the MBL phase~\cite{potter_universal_2015,dumitrescu_kosterlitz-thouless_2019}.

\section{Discussions and consequences for the MBL problem}
\label{sec:discussions}
\noindent{\it{Consequences for the recent debates---}} The overview of the problem that we propose in Fig.~\ref{fig:phasediag} for a wide range of disorder strengths, strongly  supported by a substantial body of results throughout the paper, clearly suggests that the  MBL regime is much richer than expected. Thus, while at first glance our results may seem  relatively consistent with the ``landmarks'' picture proposed by Morningstar {\it{et al.}}~\cite{morningstar_avalanches_2022}, they nevertheless deviate somewhat (as we explain below). Perhaps more importantly, we show that rare events are a key player in the very rich phenomenology of the MBL, as also recently discussed in Refs.~\cite{biroli_large-deviation_2024,colbois_statistics_2024}. 

Based on extensive numerics and an analytical ansatz, we clearly demonstrate the existence of system-wide resonances for the random-field Heisenberg chain, described in terms of long-range resonant pairs of cat states that exist for a broad range of disorder strengths, which turns out to extend far beyond the first estimates of Ref.~\cite{morningstar_avalanches_2022}. In addition we establish, based on rare events of ${\cal{O}}(1)$ two-point correlations, an upper bound for the MBL transition $h_{\rm max}\sim 20-25$, that turns out to coincide with the estimate of avalanche instability from Ref.~\cite{morningstar_avalanches_2022}.

In light of these resonant eigenstates, which incidentally are not present in the $\ell$-bit theory, it would certainly be useful to reconsider some objects that have been studied earlier in the MBL context, such as the fidelity susceptibility~\cite{sels_dynamical_2021} whose instability at strong disorder is certainly related to the existence of such resonant cat pairs~\footnote{Also discussed as Mott's pairs for Gaussian
Rosenzweig-Porter random matrix ensemble in Ref.~\cite{skvortsov_sensitivity_2022}.} but may not {\it directly} imply MBL instability, or the localization-delocalization criterion ${\cal G}$ (related to the local operator matrix elements) of Ref.~\cite{serbyn_criterion_2015} for which the probability distribution exhibits a small but noticeable increase for large ${\cal G}$ in the MBL phase.\\

\noindent{\it{Possible scenarios---}}
Returning to Fig.~\ref{fig:phasediag}, the question remains somewhat open about the nature in the the thermodynamic limit of the broad ``intermediate''  rare-event regime that we find between $h_{\rm MBL}\sim 5$, corresponding to the standard observable transitions (much less sensitive to rare events), and $h_{\rm max}\sim 20-25$, which signals the entrance in a regime where {\it{all}} eigenstates are short-range correlated and is thus presumably an upper bound for the MBL transition.  

While we are certainly better assured that this regime has a rich structure, hosting an increasing number (as $h$ decreases from $h_{\rm max}$) of resonant cat-like states, some important unanswered questions are: are these long-range correlated cat-states the main actors (i.e. are needed) for restoration of ergodicity, or a by-product of another thermalization mechanism at play? What is their required density (out of an exponential number of short-range localized ones) to witness a restoration of ergodicity in the thermodynamic limit ?

A possible scenario would be that the fat-tail regime $5\lesssim h\lesssim 10$~\cite{colbois_statistics_2024} could eventually become fully ergodic at very large scales (although impossible to observe in current numerics or experiments), due to a huge proliferation and hybridization of multiple cat-like resonances at all scales~\cite{villalonga_eigenstates_2020}. The other side of the rare-event regime, on the other hand, would likely remain stable as an MBL phase in most aspects, despite the existence of long-range resonant cat-like states, but which are not sufficiently numerous to destabilize the phase. 
Finally, we can also ask whether MBL at very strong disorder, free of rare events, is fundamentally different from MBL with very rare long-range correlated states.

\section{Summary and conclusions}

By means of large-scale, large-statistics numerical simulations, our work addressed the existence of unusual eigenstates holding anomalous strong long-range longitudinal correlations in the strong-disorder regime of the prototypical lattice model of many-body localization. In the first part of this work, we first characterize their statistics, finding that they proliferate (with their number growing with $L$ but slower that the Hilbert-space size) in a wide-(strong) disorder regime, typically $h \sim 10$ to $h \sim 20$, while we find no evidence for their probabilistic occurrence above $h \sim 20-25$. Below the value $h \sim 8-10$, their incidence is high enough that they affect average expectations values of long-range longitudinal correlators. The extreme statistics of the strongest correlator (in a disorder sample) confirms this analysis.

The second part of our paper is devoted to the microscopic content of these rare eigenstates. We make the striking observation that these eigenstates come by pairs, and that their properties are extremely well reproduced by a simple model of toy cat states. The reproduced properties are: the strong similarity (as measured by a low Kullback-Leibler divergence) between the two states in the pair, the fact that the two states of the pair can be mapped one to another by simply applying $\sigma_i^z$
($\Sigma$ test) and of course the strong defining connected longitudinal correlations. Further, we find that these long-range cat states are most of the time nearest-neighbors in the energy spectrum, separated by an energy gap which is statistically smaller than the many-body level spacing.
All these elements provide a proof that these eigenstates pairs are nothing but long-range resonances which have been extensively discussed and argued to be the precursors of thermalization inside the MBL regime~\cite{gopalakrishnan_low_2015,villalonga_eigenstates_2020,garratt_local_2021,garratt_resonant_2022,morningstar_avalanches_2022,crowley_constructive_2022,Ha_many-body_2023}. While some previous works postulated the existence (and importance) of cat states as incarnations of many-body resonances inside the MBL phase, we believe this is the first complete and extended demonstration of (a category of) cat eigenstates (those carrying long-range strong correlations) as long-range many-body resonances for a microscopic model hosting a MBL phase. Previous numerical work in the quest for MBR indeed either tested {\it consequences of} (or directly studied) theoretical models of MBR ~\cite{gopalakrishnan_low_2015,geraedts_many-body_2016,kjall_many-body_2018,villalonga_eigenstates_2020,crowley_constructive_2022,morningstar_avalanches_2022}, with the exception of Ref.~\cite{Ha_many-body_2023}. Here we instead provide a concrete filtering procedure for selecting these rare eigenstates, which are then {\it a posteriori} found to be many-body resonances of cat states type.\\

Our results provide a number of ways forward to improve our understanding of the MBL problem in the strongly debated intermediate to strong disorder regime $h>5$. 
First, they offer a concrete basis for improving models~\cite{garratt_local_2021,crowley_constructive_2022,garratt_resonant_2022}  of many-body resonances in the most studied MBL Hamiltonian, for instance allowing one to test directly their consequences. The intriguing scaling of the number of fluctuating spins, together with the cat-states magnetization profiles that we report, do not immediately match with the naive perturbative expectation of $L/2$ spin flips to create cat states resonating at distance $L/2$. Further, the filtering process for identifying the cat-states MBR could also be used in the core of renormalization group treatments~\cite{pekker_hilbert-glass_2014, potter_universal_2015,khemani_obtaining_2016} of the MBL problem. Finally, the sparsity of these cat-states at (very) strong disorder, and their increased occurrence as disorder decreases, confirm the validity of rare-events analysis of the MBL regime~\cite{biroli_large-deviation_2024}.\\

For concreteness, we focused our analysis only on system-wide (in our case the maximal $L/2$ distance) correlations, as they are expected to render the MBL phase unstable, thus providing an upper bound on MBL stability (which our analysis suggests to be $h_{\rm max} \sim 20-25$ for the random-field Heisenberg chain). There are of course resonances on {\it smaller} length-scales (at short fixed distances, or at other extensive scales such as $L/3$, $L/4$, etc.), for which a similar analysis could also be performed. We expect that the toy cat states should also provide a good modelization of such shorter-range resonances (indeed there is no dependence on the distance between resonating sites in the toy cat states). The only difference would be in the energy scale separating the cat states (the shorter-range resonances will be more distant in the energy spectrum). Another intrinsic property of the random-field Heisenberg chain is that it conserves magnetization, which is not necessary for MBL to occur. It would be extremely appealing to extend our approach and the use of the proposed numerical probes for catness (such as low KL, the $\Sigma$ test, modelization with toy cat states) in models of MBL with no U($1$) conservation law~\cite{pekker_hilbert-glass_2014,huse_localization_2014,kjall_many-body_2014,zhang_floquet_2016,PhysRevLett.126.100604,laflorencie_topological_2022,PhysRevB.105.144205,PhysRevResearch.6.L032045}.\\
\\

\acknowledgments
We thank G. Alfaro, M. Tarzia, A. Padhan, K. Pawlik and J. Zakrzewski for discussions and collaborations on related projects. We also acknowledge D. Huse, I. Khaymovitch, and V. Kravtsov for comments on the manuscript. This work has been partly supported by the EUR grant NanoX No. ANR-17-EURE0009 in the framework of the ”Programme des Investissements d’Avenir”, is part of HQI initiative (www.hqi.fr) and is supported by France 2030 under the French National Research Agency award number ANR- 22-PNCQ-0002. This work is also supported by the ANR research grant ManyBodyNet No. ANR-24-
CE30- 5851, and also benefited from the support of the Fondation Simone et Cino Del Duca. We acknowledge the use of HPC resources from CALMIP (grants 2024-P0677) and GENCI (project A0170500225), as well as of the PETSc~\cite{petsc-user-ref,petsc-efficient}, SLEPc~\cite{slepc-toms,slepc-users-manual}, and MUMPS~\cite{MUMPS1,MUMPS2} sparse linear algebra libraries.

\newpage
\setcounter{section}{0}
\setcounter{secnumdepth}{3}
\setcounter{figure}{0}
\renewcommand\thesection{S\arabic{section}}
\renewcommand\thefigure{S\arabic{figure}}
\renewcommand\theequation{S\arabic{equation}}

\appendix
\begin{center}
    \bfseries\Large Supplementary Material
\end{center}
These appendices support several points which are only briefly mentioned in the main text. Appendix~\ref{app:twopoint} focuses on two point-functions, especially at large distances, while Appendix~\ref{app:LRCCs} focuses on both the toy and the actual cat states properties. 
\section{Strong systemwide two-point functions}
\label{app:twopoint}
\subsection{Instability regime}

Here, we provide a brief reminder of the detection of an instability regime discussed extensively in Ref.~\cite{colbois_statistics_2024}, taking advantage of our new data up to extremely strong disorder. Unlike that previous work, we here use the full many-body spectrum. As a consequence, we have access to much larger statistics, but are limited to smaller system sizes. 

\begin{figure}[h!]
    \centering
    \includegraphics[width=0.65\columnwidth]{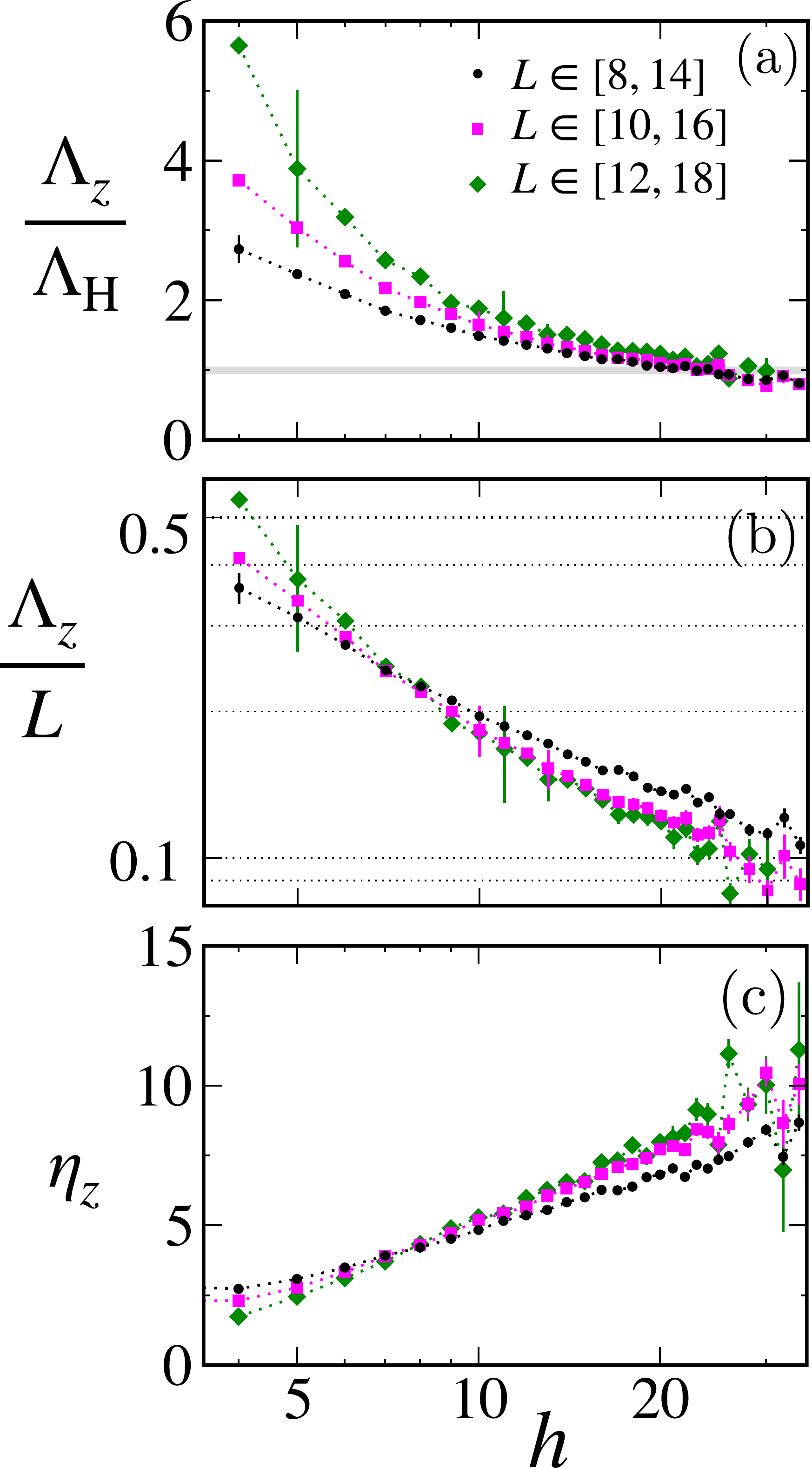}
    \caption{Parameters controlling the decay of average number of eigenstates per sample $N^{z}_{\star}$ having strong systemwide correlations for $C_{\star} = 0.1$ (see also caption of Fig.~\ref{fig:Nz}). (a) $\Lambda_z/\Lambda_H$ becomes smaller than 1 for $h \sim 20-25$. (b) The rescaling of $\Lambda_z$ with the average system size of the fit window ($L = 11, 13, 15$) is consistent with an instability regime gradually setting in below $h\sim 8-10$. (c) This is consistent with the results for the power-law exponent $\eta_z$. }
    \label{fig:Lambda_Czz_and_Eta}
\end{figure}

As discussed in Sec.~\ref{sec:stats}, we focus on two related observables: the weight $W_{\star}^{z}$ Eq.~\ref{eq:Wzstar} of eigenstates with anomalously large mid-chain longitudinal correlations, and the related average number $N_{\star}^{z}$ of such eigenstates per sample Eq.~\ref{eq:Nzstar}, where the $\star$ symbol refers to the dependence on the criterion $C_{\star}$. 
Fig.~\ref{fig:Lambda_Czz_and_Eta} considers the system size dependence based on two types of fits for the weight: power-law Eq.~\ref{eq:Wzstar_pl}, controlled by $\eta_z$, or exponential Eq.~\ref{eq:Wzstar_exp}, controlled by $\Lambda_z$. Their comparison for sliding fits over four system sizes highlights three regimes, for $C^{zz}_{L/2} \geq C_{\star} = 0.1$. 
The behavior of $\Lambda_z$, compared to the natural length $\Lambda_H$ associated with the many-body spectral gap in panel (a), is consistent with the inset of Fig.~\ref{fig:Nz} at $C_{\star} = 0.2$: at very strong disorder $h \gtrsim 20-25$, $\Lambda_z/\Lambda_H$ is well converged and below unity, suggesting a complete absence of rare events in the thermodynamic limit (scenario (ii) in Sec.~\ref{sec:scaling}). 

At weaker disorder ($h \lesssim 20)$, we ask whether we can distinguish between possibilities (i) and (iii). Namely, following Ref.~\cite{colbois_statistics_2024}, we seek to distinguish (i) the existence of a well-converged $\Lambda_z > \Lambda_H$ from (iii) a power-law dependence of the weight $W_{\star}^{z}$ on the system size, implying a growing effective $\Lambda_z$ and a converged $\eta_z$. Panels (a-b-c) of Fig.~\ref{fig:Lambda_Czz_and_Eta} provide a consistent picture. While it is difficult to pinpoint from panel (a) exactly at which disorder strength $\Lambda_z$ will converge to a finite value, the results in panels (b-c) show beyond any doubt that the decay $W_{\star}^{z}(L)$ is much slower than exponential for $h \lesssim 8-10$ - it may even be slower than power-law. This is consistent with the result of Ref.~\cite{colbois_statistics_2024}, where a much smaller threshold $C_{\star} = 0.01$ was used. Although limited system sizes do not allow clear conclusions, the results for the larger system sizes could potentially be consistent with a power-law regime extending to slightly stronger disorder strength. This highlights the challenge of getting both large enough systems and sufficient statistics of the rare events of strong correlations to conclude on the distinction between (i) and (iii).

\subsection{Quantum mutual information}
\label{app:qmi}
\subsubsection{Reduced density matrix}
The reduced density matrix  $\rho_{ij}$ of  two spins (at sites $i$ and $j$) can be expanded~\cite{latorre_ground_2004} using the Pauli matrices, 
\be
\rho_{ij}=\frac{1}{4}\sum\nolimits_{\alpha,\beta=1,x,y,z}\langle\sigma_{i}^{\alpha}\sigma_{j}^{\beta}\rangle \sigma_{i}^{\alpha}\sigma_{j}^{\beta},
\ee
where $\alpha,\beta=1$ corresponds to the $2\times 2$ identity matrix. For the model Eq.~\ref{eq:RFHCM}, the magnetization conservation yields non-zero contributions only for the $11$, $xx$, $yy$, $zz$, $1z$, $z1$ combinations, giving

\[
\rho_{ij}=\frac{1}{4}\left(
\begin{array}{cccc}
1  & 0  &0  &0 \\
 0 &  1 &  0 & 0\\
  0& 0  & 1  & 0\\
 0 &  0 &  0 & 1\\
\end{array}
\right)
\]
\[
+\frac{1}{2}\langle\sigma_{i}^{x}\sigma_{j}^{x}\rangle\left(
\begin{array}{cccc}
0 &  0 & 0 & 0\\
  0&  0 & 1  & 0\\
  0&  1 & 0  & 0\\
  0& 0  &0   &0 \\
\end{array}
\right)
+\frac{1}{4}\langle\sigma_{i}^{z}\sigma_{j}^{z}\rangle\left(
\begin{array}{cccc}
1 &  0 & 0 & 0\\
  0&  -1 & 0  & 0\\
  0&  0 & -1  & 0\\
  0& 0  &0   &1 \\
\end{array}
\right)
\]
\[
+\frac{1}{4}\langle\sigma_{i}^{z}\rangle\left(
\begin{array}{cccc}
1 &  0 & 0 & 0\\
  0&  1 & 0  & 0\\
  0&  0 & -1  & 0\\
  0& 0  &0   &-1 \\
\end{array}
\right)
+\frac{1}{4}\langle\sigma_{j}^{z}\rangle\left(
\begin{array}{cccc}
1 &  0 & 0 & 0\\
  0&  -1 & 0  & 0\\
  0&  0 & 1  & 0\\
  0& 0  &0   &-1 \\
\end{array}
\right).
\]
If one defines 
\bea
m_i&=&\langle\sigma_{i}^{z}\rangle\\
m_j&=&\langle\sigma_{j}^{z}\rangle\\
c^{zz}_{ij}&=&\langle\sigma_{i}^{z}\sigma_{j}^{z}\rangle\\
c^{xx}_{ij}&=&\langle\sigma_{i}^{x}\sigma_{j}^{x}\rangle,
\eea
the reduced density matrix takes the simple block-diagonal form
\[
\rho_{ij}=\frac{1}{4}\left(
\begin{array}{cccc}
1  & 0  &0  &0 \\
 0 &  1 &  0 & 0\\
  0& 0  & 1  & 0\\
 0 &  0 &  0 & 1\\
\end{array}
\right)
\]
\[
+\left(
\begin{array}{cccc}
\frac{m_i+m_j+c^{zz}_{ij}}{4}  & 0  &0  &0 \\
 0 &  \frac{m_i-m_j-c^{zz}_{ij}}{4} &  \frac{c^{xx}_{ij}}{2} & 0\\
  0& \frac{c^{xx}_{ij}}{2}  & -\frac{m_i-m_j+c^{zz}_{ij}}{4} & 0\\
 0 &  0 &  0 & -\frac{m_i+m_j-c^{zz}_{ij}}{4}\\
\end{array}
\right),
\]
from which the 4 eigenvalues are easy to get:
\bea
\label{eq:lambdas}
\lambda_1&=&{(1+m_i+m_j+c^{zz}_{ij})}/{4}\nonumber\\
\lambda_2&=&{(1-m_i-m_j+c_{ij}^{zz})}/{4}\\
\lambda_{3,4}&=&\frac{1}{4}\left(1-c_{ij}^{zz}\pm\sqrt{(m_i-m_j)^2+(2c_{ij}^{xx})^2}\right).\nonumber
\eea
\subsubsection{Quantum mutual information}
The quantum mutual information
\be
{\rm QMI}_{ij}=S^{(1)}_{i}+S^{(1)}_{j}-S^{(2)}_{ij},
\ee
involves the entanglement entropies (EEs)  of the single sites $S_i$ and $S_j$, and the EE of the pair $S_{ij}$.
In the model at hand Eq.~\ref{eq:RFHCM}, the single-site reduced density matrix is diagonal with entries $\frac{1}{2}\left(1\pm\langle \sigma^z\rangle\right)$, which gives for the single-site EE
\bea
S_i&=&\ln 2-\frac{1}{2}\left(1-m_i\right)\ln\left(1-m_i\right)\nonumber\\
&-&\frac{1}{2}\left(1+m_i\right)\ln\left(1+m_i\right).
\eea
The two-site EE 
\be
\label{eq:2sEE}
S_{ij}=-\sum_{k=1}^{4}\lambda_k\ln\lambda_k,
\ee
can be simplified in some limits, using the expressions of the $\lambda$s given in Eq.~\eqref{eq:lambdas}
\subsubsection{QMI for toy cat sates}
For the general ansatz description of cat pairs
\begin{equation*}
    \begin{array}{l}
{\ket{{\Phi_{2p}^{+}}}}=\left(\cos\left(\frac{\theta}{2}\right){\ket{\varphi_{2p}}}
+\sin\left(\frac{\theta}{2}\right){\overline{{\ket{\varphi_{2p}}}}}\right)\otimes {\ket{\varphi_{L-2p}}}\\
{\ket{{\Phi_{2p}^{-}}}}=\left(\sin\left(\frac{\theta}{2}\right){\ket{\varphi_{2p}}}
-\cos\left(\frac{\theta}{2}\right){\overline{{\ket{\varphi_{2p}}}}}\right)\otimes {\ket{\varphi_{L-2p}}},
\end{array}
\end{equation*}
where the ${\ket{\varphi_{2p}}}$ are $2p$ site product states of zero total magnetization, one has to consider the three possible situations:\\

(i) If $p=1$, this is a Bell state of the form $\Psi$ for which $m_i=-m_j=\pm \cos\theta$, $c_{ij}^{zz}=-1$, and $c_{ij}^{xx}=-\sin\theta$.\\

(ii) If $p>1$ and correlations are {\it{antiferromagnetic}}, $m_i=-m_j=\pm \cos\theta$, $c_{ij}^{zz}=-1$, and $c_{ij}^{xx}=0$.\\

(iii) If $p>1$ and correlations are {\it{ferromagnetic}}, $m_i=m_j=\pm \cos\theta$, $c_{ij}^{zz}=1$, and $c_{ij}^{xx}=0$.\\

In all the three cases, the sum of the two single-site EEs is given by
\bea
\label{eq:SiplusSj}
S_i+S_j&=&\ln 4-\left(1-\cos\theta\right)\ln\left(1-\cos\theta\right)\nonumber\\
&-&\left(1+\cos\theta\right)\ln\left(1+\cos\theta\right).
\eea
On the other hand, the eigenvalues $\lambda_i$, Eq.~\eqref{eq:lambdas} take distinct forms. For the case (i), we get $\lambda_{1,2,4}=0$ and $\lambda_3=1$, thus yielding for the two-site EE Eq.~\eqref{eq:2sEE} $S_{ij}=0$, and therefore ${\rm QMI}=S_i+S_j$, as given by Eq.~\eqref{eq:SiplusSj}. There is a symmetry between case (ii), which gives  $\lambda_{1,2}=0$, and $\lambda_{3,4}=(1\pm\cos\theta)/2$, and case (iii) for which $\lambda_{1,2}=(1\pm\cos\theta)/2$, and $\lambda_{3,4}=0$, thus yielding for (ii-iii) $S_{ij}=(S_i+S_j)/2$, and thus ${\rm QMI}=(S_i+S_j)/2$.

We can summarize the results in a single formula
\be
{\mathrm{QMI}}=Q_{p}\left[\ln \left(\frac{2}{\sin\theta}\right) -\frac{\cos\theta}{2}\ln\left(\frac{1+\cos\theta}{1-\cos\theta}\right)\right],
\ee
where the prefactor simply encodes the type of cat state, $Q_p=2$ for $p=1$ and $Q_p=1$ for $p>1$.

\subsection{Edge effects}
\label{app:edges}
\begin{figure}[t!]
    \centering
    \includegraphics[width=\columnwidth,clip]{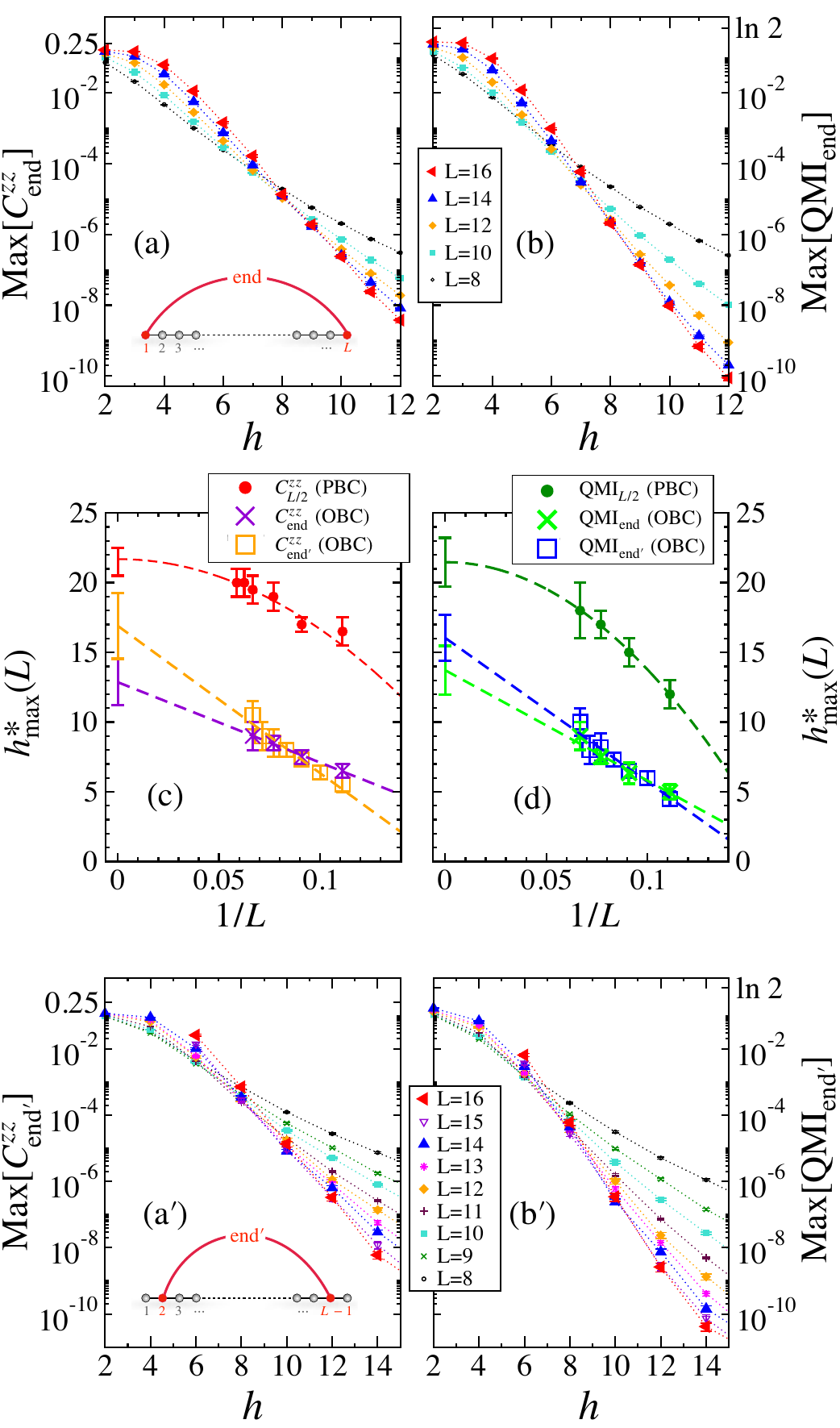}
    \caption{Boundary effects on the large long-distance correlations providing the ``systemwide resonances'' landmark ${\cal{L}}_{\rm swr}$ based on the maximal QMI. Left: longitudinal correlations. Right : quantum mutual information Eq.~\ref{eq:QMI}. (a-b) : reproducing the results of Ref.~\cite{morningstar_avalanches_2022} and comparing to the longitudinal correlations result, $h_{\rm{max}}(L) \in  7-9$. (a'-b') The results are modified significantly when considering correlations between sites $2$ and $L-1$ instead, $h_{\rm{max}}(L)\in [6-11]$. (c-d) The drifts of the crossings between neighbouring even system sizes are shown as a function of $1/L$, with $L$ indicating the largest of the two sizes, and extrapolate to widely different $h_{\rm{max}}$. }
\label{fig:OBCs} 
\end{figure}

\begin{figure}[t!]
    \centering
   \includegraphics[width=.6\columnwidth]{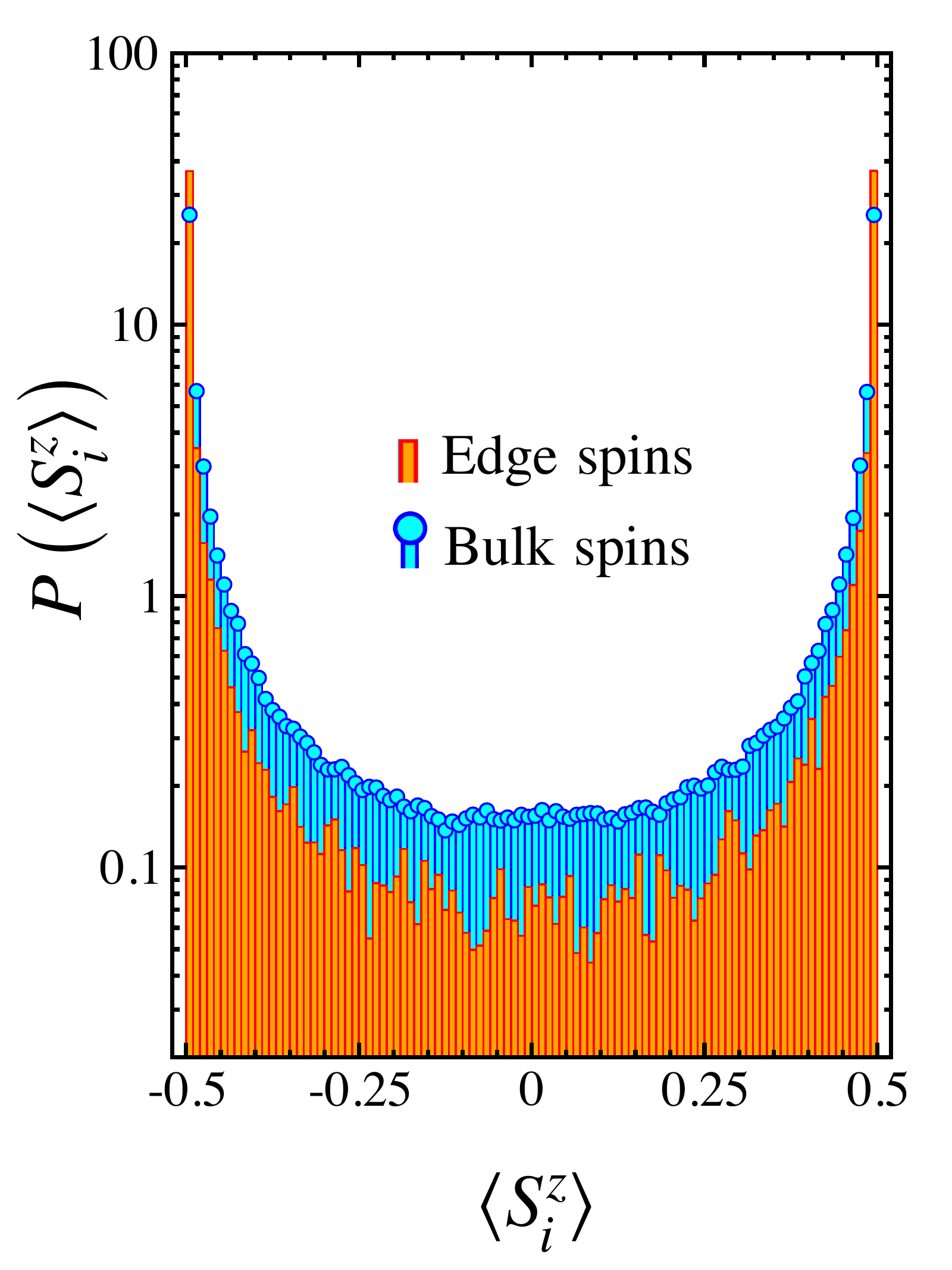}
    \caption{For open chains, the edge spins are more localized than the bulk ones. Histograms of the local magnetizations collected over $10^3$ samples and all eigenstates for $L=16$ OBC chains at $h=10$.}
    \label{fig:OBC_Bulk_vs_Edge}
\end{figure}

We provide evidence supporting our claim of Sec.~\ref{sec:boundary} regarding the boundary effects. First, we confirm the results obtained in Ref.~\cite{morningstar_avalanches_2022} for the system-wide resonances landmark ${\cal{L}}_{\rm swr}$, see panels (a-b) of Fig.~\ref{fig:OBCs}. 
As claimed in the main text, Fig.~\ref{fig:OBC_Bulk_vs_Edge} provides clear evidence that the two end spins are effectively more localized than the bulk spins. We proceed to show that this has important consequences on the landmarks given by the maximal longitudinal correlations or the maximal QMI. Indeed, if we take the two spins which are one site away from the edge of the chains (which we dub {\it{end'}}, see panels (a') and (b') of Fig.~\ref{fig:OBCs}), the critical value (as obtained by the crossings) occurs at larger disorder. Furthermore, its drift with system size gets more pronounced. This effect is emphasized in panels (c) and (d), where the values of this landmark for OBC (end and end') are compared to the one obtained with PBCs. In the latter case, the landmark ${\cal{L}}_{\rm swr}$ gets pushed all the way to $h \sim 20$, as discussed in the main text.

\section{Long-range correlated cat states}
\label{app:LRCCs}

This appendix provides additional informations on the toy cat states including examples and detailed calculations.

\subsection{Example for $L = 20$ and $h = 9$}
\label{app:examplemore}
In Fig.~\ref{fig:sample_L20_h9}, we presented an example of a strongly correlated state (Sample 2) which guided our construction of toy cat states. However, we later focused on other examples in Sec.~\ref{sec:examples} when testing out catness measures.
For the sake of completeness, Table~\ref{tab:samples2} provides the results associated with Sample 2. They are characteristic of a cat state pair of type $\Phi_{2p}$ with $p > 1$ (Eq.~\ref{eq:Phi2p}, Table~\ref{tab:catstates}), as highlighted by the very small value of the transverse correlations and by the QMI a bit smaller than $Q_2 = \ln(2)$.  The ``$\Sigma$-test'' reveals a less perfect cat pair than Sample 3 in Table~\ref{tab:samples}. 

\begin{table}[h!]
    \centering
    \begin{tabular}{l|c|c|}
& \multicolumn{2}{c|}{{\rm{Sample}} 2} \\
\hline
$\epsilon$& 0.5000330 & 0.5000366\\
\hline  
        $\Delta\epsilon$ & \multicolumn{2}{c|}{$3.62\times 10^{-6}$} \\
         \hline
         $4C_{L/2}^{zz}$& 0.91371 & 0.92705 \\
         \hline
         $4C_{L/2}^{xx}$& $-3.3 \times 10^{-6}$ &$-2.7 \times 10^{-6}$ \\
         \hline
         $2\langle S_i^z \rangle$& -0.01562 & 0.05118   \\
         $2\langle S_{i+L/2}^z \rangle$& -0.0272  & 0.05750 \\
                  \hline
         QMI & 0.5161 & 0.5401 \\
                  \hline
         KL & \multicolumn{2}{c|}{0.01718} \\
         \hline
         $\Sigma_i^\pm$ & \multicolumn{2}{c|}{0.94201} \\
         $\Sigma_{i+L/2}^\pm$ & \multicolumn{2}{c|}{0.97691} \\
         \hline
         $4C^{zz}_{L/2}-\Sigma_i^\pm\Sigma_j^\pm$& -0.00012 & 0.00006 \\
         \hline
    \end{tabular}
    \caption{Sample 2 from Fig.~\ref{fig:sample_L20_h9} ($L = 20, h = 9$), where $i=4, j=i+L/2=14$ are the sites with a strong longitudinal correlation.}
    \label{tab:samples2}
\end{table}

\subsection{KL for cat states}
\label{app:KLcat}
In the main text Sec.~\ref{sec:Catness}, we discussed properties of potential cat states under the light of toy cat states. There, we mentioned results for the average and typical KL in a simplified picture where the distribution of $\theta$ is modeled as uniform in $[\theta^{\star}, \pi/2]$. Here we provide a few details. 

The average KL is given by
\begin{eqnarray}
{\overline{{\rm KL}_{\rm CS}^{[\pm]}(\theta_\star)}}&=&\int_{\theta_\star}^{\pi/2} \frac{{\rm KL}_{\rm CS}^{[\pm]}(\theta)}{\pi/2-\theta_\star} {\rm{d}}\theta \label{eq:avgKLcs} \\
&=&\int_{-\delta_\star}^{0} \frac{{\rm KL}_{\rm CS}^{[\pm]}(\delta)}{\delta_\star}{\rm{d}}\delta\nonumber \\
&\approx&\frac{2}{3}\delta_\star^2 -\frac{1}{5}\delta_\star^4\nonumber \\
&\approx&\frac{2}{3}\left[\frac{\pi}{2}-\arcsin{\left(2\sqrt{C_\star}\right)}\right]^2\nonumber.
\end{eqnarray}
In the first two steps we have replaced ${\rm KL}^{\pm}_{\rm CS}$ by its expansion around $\delta = \pi/2 -\theta  =0$  (Eq.~\eqref{eq:KL_expansion}) before integrating. The last equality comes from replacing $\delta_{\star}$ in the 2nd order term by its expression in terms of the criterion $C_\star$. 
\begin{eqnarray}
\exp\left({\overline{\ln {{\rm{KL}}}_{\rm CS}}}\right) &=&\exp\left(\int \ln\left[{\rm KL}^\pm(\theta)\right]P(\theta){\rm{d}}\theta\right) \label{eq:typKLcs} \\
&\approx & \frac{2}{{\rm{e}}^2}\left(\frac{\pi}{2}-\theta_{\star}\right)^2\nonumber \\
&\approx & \frac{2}{{\rm{e}}^2}\left[\frac{\pi}{2}-\arcsin{\left(2\sqrt{C_\star}\right)}\right]^2\nonumber.
\end{eqnarray}

 \begin{figure}[t!]
    \centering
    \includegraphics[width=.825\columnwidth]{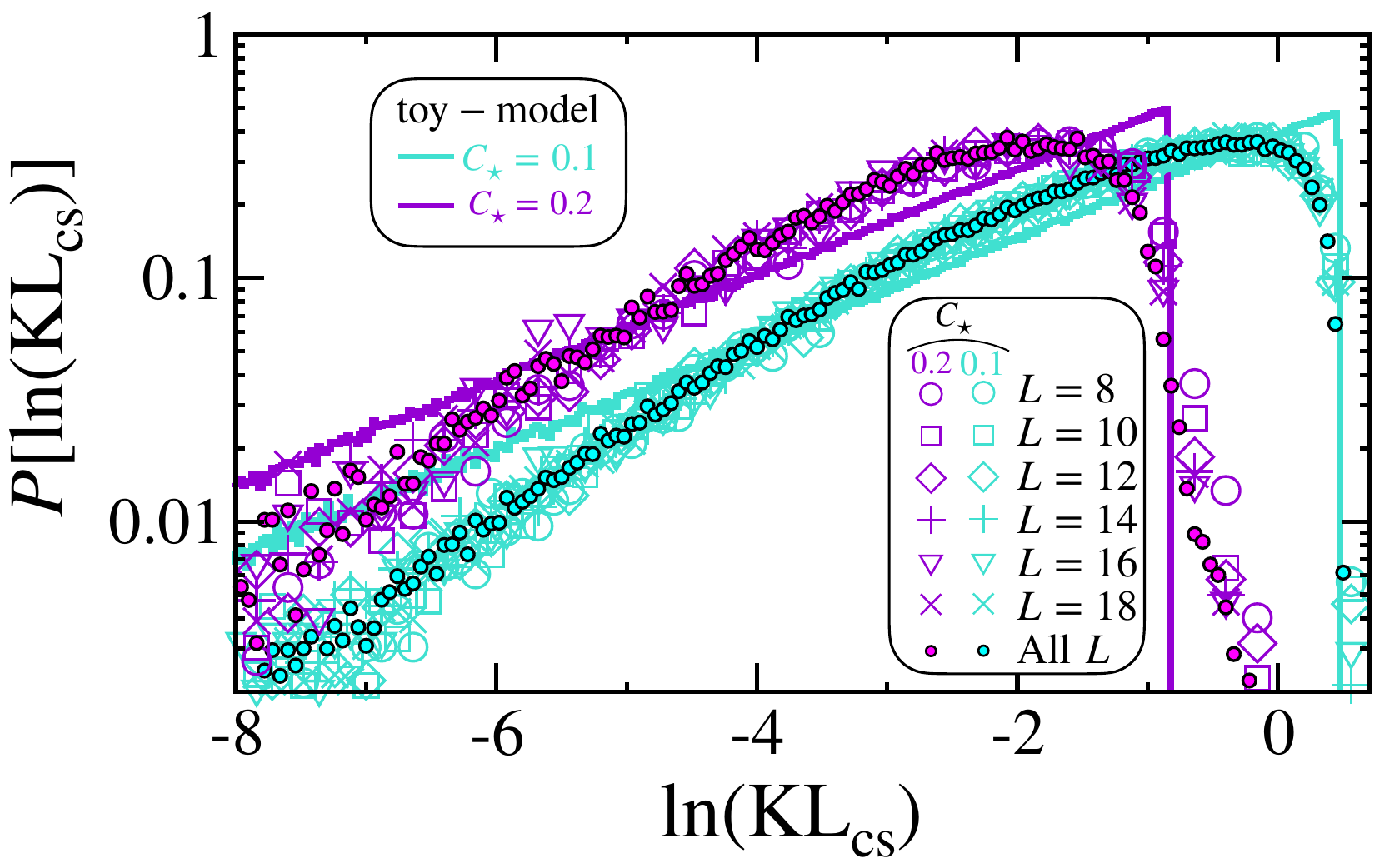}
    \caption{Distribution of the Kullback-Leibler divergence Eq.~\eqref{eq:KL_Def} for the cat states, selected as in Sec.~\ref{sec:selectionalgorithm}. The results for toy states are obtained  assuming a uniform distribution of $\theta$ between $\theta_{\star}$ and $\pi/2$ in the expression Eq.~\eqref{eq:KLCat}.}
    \label{fig:HistKLCats}
\end{figure}

We show in Fig.~\ref{fig:HistKLCats} the distributions $P(\ln {\rm{KL}_{\rm cs}})$ for two values of $C_\star$.
As we have previously seen that there is no $h$-dependence for $h\ge 10$ (see Fig.~\ref{fig:KLCats_vs_h}), all values of $h\ge 10$ are treated equally to build the histograms. This plot further demonstrates the lack of $L$ dependence, and the reasonably good agreement between  the histogram of ED data and the  toy model, made from Eq.~\eqref{eq:KL_expansion} and a simple box distribution for $\delta\in [-\delta_\star ; 0]$, with $\delta_\star=\arccos\left(2\sqrt{C_\star}\right)$. As expected, not all the details of the distributions are captured, but several very important features are well explained by the model, such as the strong skewness, the exponential tail, and the overall values are quantitatively quite good.

\subsection{Histogram of ${\mathsf{N}}_0$ for standard MBL eigenstates}
 \begin{figure}[b]
     \centering
    \includegraphics[width=\columnwidth]{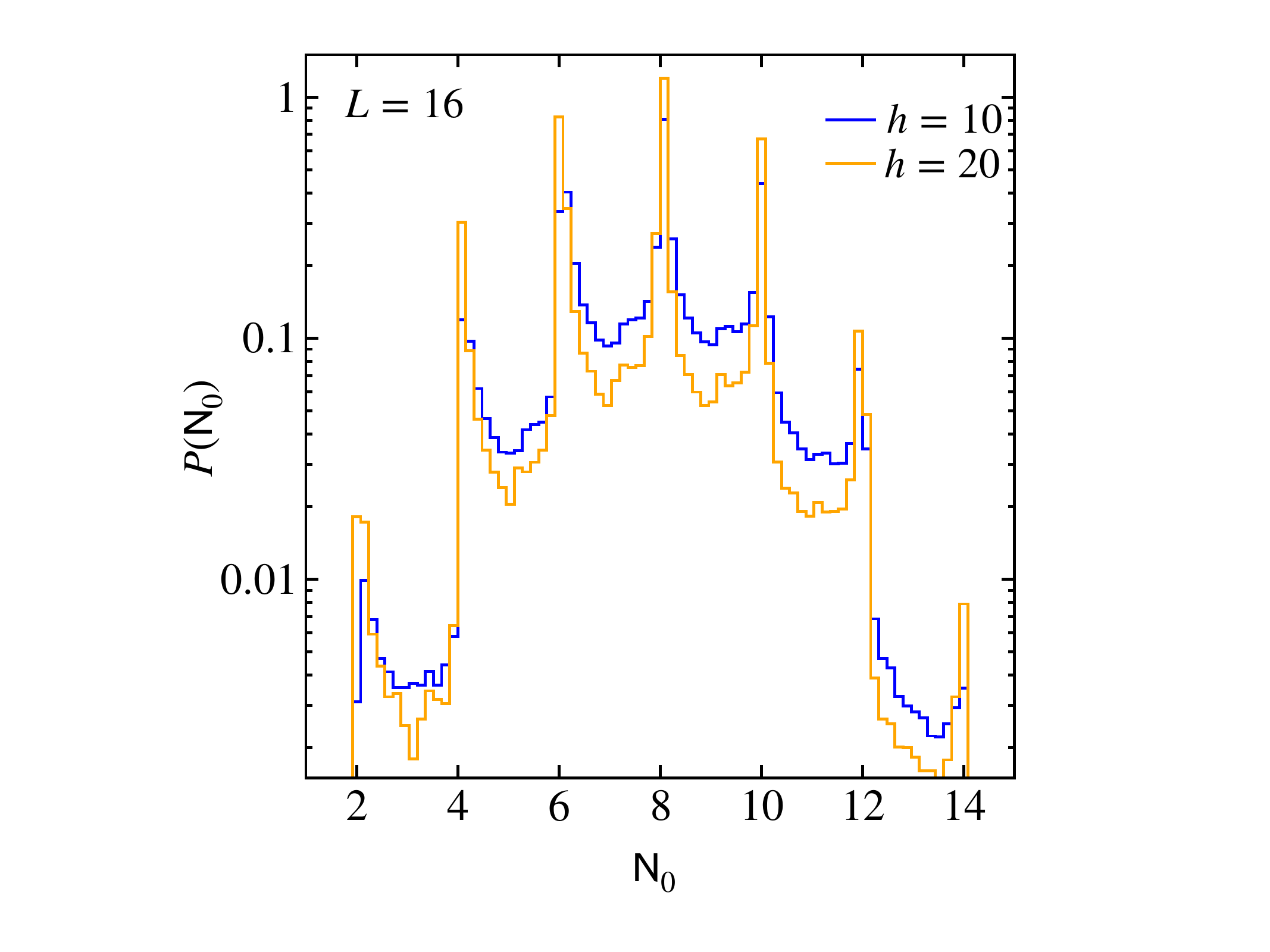}
    \caption{Same as Fig.~\ref{fig:HistoN0} of the main text, but here the data ($L=16$ and $h=10,\,20$) are not restricted by the cat-state filtering: {\it{all}} adjacent eigenstates are used in the computation of ${\mathsf{N}}_0$, Eq.~\eqref{eq:N0}, and the the histogram, dominated by  $\ell$-bits type MBL eigenstates is strongly peaked for even values with an overall gaussian enveloppe centered around 
${\mathsf{N}}_0= L/2$.}
    \label{fig:Hist0_N0_MBL}
 \end{figure}
\label{sec:PN0MBL}
Here we provide an additional plot that shows how the distribution $P({\mathsf{N}}_0)$ behaves for standard MBL eigenstates. Introduced in Sec.~\ref{sec:CatMicroscopy} to analyze the number of resonant spins, and discussed in Fig.~\ref{fig:HistoN0} for cat states, it was argued that ${\mathsf{N}}_0$ provides a good proxy for the number of fluctuating sites within cat pairs. Here we show in Fig.~\ref{fig:Hist0_N0_MBL} that by simply taking {\it{all}} adjacent eigenstates for  ${\ket{\Phi^\pm}}$ in the computation of ${\mathsf{N}}_0$ defined in Eq.~\eqref{eq:N0} (that is, without applying the cat-state filtering algorithm), the histogram is dominated by MBL eigenstates with localized bits. As anticipated, $P(\mathsf{N}_0)$ is strongly peaked at even values with an overall gaussian envelope centered around 
${\mathsf{N}}_0= L/2$. In this case, ${\mathsf{N}}_0$ does not have the same interpretation as for cat-states.

\subsection{Decay of cat state gaps}
\label{app:spectro}
 \begin{figure*}[ht]
     \centering
    \includegraphics[width=1.65\columnwidth]{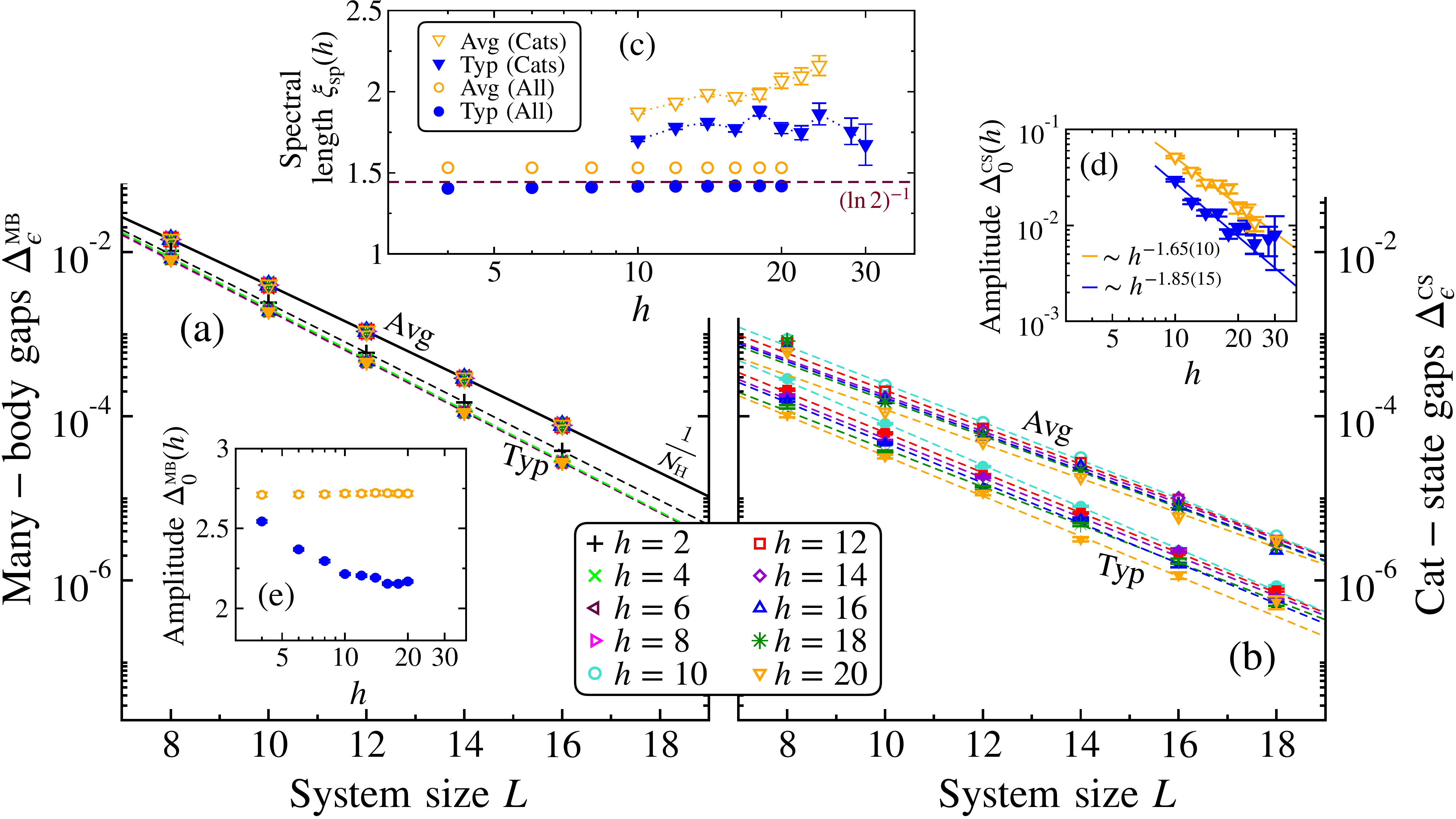}
    \caption{Comparison between the typical and average many-body gaps (a) and the typical and average cat gaps (b), for all disorder strengths and available system sizes. The insets show the parameters for fits following Eq.~\eqref{eq:Deltaepsilon}:  (c) spectral length, (d-e) amplitudes (only for the cat states).}
    \label{fig:CatGaps}
 \end{figure*}

Here, we provide further information with respect to the cat states spectroscopy discussed in Sec.~\ref{sec:CatSpectroscopy}. 
Fig.~\ref{fig:CatGaps} shows the results for the finite-size scalings of average and typical cat-state gaps, accompanied by a detailed comparison with the scaling of natural many-body level spacings. The numerical data for average and typical gaps are very well described by the following exponential decay
\be
\Delta_{\epsilon}(h,L)= \Delta_{0}\exp\left(-\frac{L}{\xi_{\rm sp}(h)}\right).
\label{eq:Deltaepsilon}
\ee
The many-body gap is very well described by a spectral decay length $\xi_{\rm sp}\approx (\ln 2)^{-1}$, see Fig.~\ref{fig:CatGaps} (a) for both average and typical decays. The average (open symbols) almost perfectly follows ${\overline{\Delta_{\epsilon}^{_{\rm MB}}}}\approx \frac{1}{{\cal{N}}_{\rm H}}$, for all values of $h$. The typical many-body gap (filled symbols), is always smaller than the average, but also decays exponentially with $\xi_{\rm sp}\approx (\ln 2)^{-1}$, and with an amplitude $\Delta_{0}$ which shows a (weak) $h$-dependence, but remains ${\cal{O}}(1)$, see panel (e).\\

The cat-state gaps, as already seen in the histograms (Fig.~\ref{fig:HistoCatGaps}), are significant smaller than their natural many-body counterparts, and that is true for both average or typical values, as clearly visible in Fig.~\ref{fig:CatGaps} (b). However, the exponential decays of the data, following Eq.~\eqref{eq:Deltaepsilon}, are best described by an almost $h$-independent spectral length $\xi_{\rm sp}$ which is slightly larger than the natural length $(\ln 2)^{-1}$, as shown in Figure.~\ref{fig:CatGaps} (c), but with a prefactor $\Delta_0(h)$ which strongly depends on the disorder strength, see Fig.~\ref{fig:CatGaps} (d).

While both the typical and average cat gaps a priori show a slower exponential decay than the natural level spacing, this effect is largely offset by the much smaller amplitude of the cat gaps, at least on the available system sizes. However, we are not in a position to conclude whether at very large sizes the natural gap will become smaller than the cat states gap.

\newpage
\bibliography{CATS_MBL}

\end{document}